\documentstyle[prd,aps,epsf,floats,amsfonts,amssymb,amsmath]{revtex}
\newcommand{\bea}{\begin{eqnarray}} \newcommand{\eea}{\end{eqnarray}}
\newcommand{\ba}{\begin{array}} \newcommand{\ea}{\end{array}}
\newcommand{\bit}{\begin{itemize}} \newcommand{\eit}{\end{itemize}}
\newcommand{\ben}{\begin{enumerate}} \newcommand{\een}{\end{enumerate}}
 \newcommand{\lab}{\label}
\newcommand{\lan}{\langle} 
\newcommand{\lf}{\left} \newcommand{\lrar}{\leftrightarrow}
 \newcommand{\noi}{\noindent}
\newcommand{\non}{\nonumber} 
\newcommand{\pa}{\partial} \newcommand{\ran}{\rangle}
\newcommand{\rar}{\rightarrow} 
\newcommand{\ri}{\right} \newcommand{\ti}{\tilde}

 \newcommand{\bt}{\beta}
\newcommand{\ga}{\gamma} \newcommand{\Ga}{\Gamma}
\newcommand{\de}{\delta} \newcommand{\De}{\Delta}
 
\newcommand{\te}{\theta} 
 \newcommand{\la}{\lambda}
\newcommand{\La}{\Lambda}

\newcommand{\ident}{1\hspace{-1mm}{\rm I}}
\newcommand{\cali}[1]{\cal #1}

\begin{document}

\typeout{--- Title page start ---}

\title{
Topological defects as inhomogeneous condensates  in Quantum Field
Theory:\\[2mm] Kinks in ($1+1$) dimensional $\la \psi^4$ theory}

\author{Massimo Blasone${}^{\sharp}$ and Petr Jizba${}^\flat$}

\address{
$ $ \\
${}^{\sharp}$   Blackett  Laboratory, Imperial College, Prince
Consort Road, London  SW7 2BZ, U.K. \\
and Dipartimento di  Fisica
and INFN, Universit\`a di Salerno, I-84100 Salerno, Italy
\\ [1mm] ${}^{\flat}$ Institute of Theoretical Physics, University of Tsukuba,
Ibaraki 305-8571, Japan
\\[1mm] E-mails:  m.blasone@ic.ac.uk, petr@ph.tsukuba.ac.jp
\\[2mm]}

\maketitle

\begin{abstract}
We study topological defects as inhomogeneous (localized)
condensates of particles in Quantum Field Theory. In the framework
of the Closed--Time--Path formalism, we consider explicitly a
$(1+1)$ dimensional $\la \psi^4$ model and construct the
Heisenberg picture field operator $\psi$ in the presence of kinks.
We show how the classical kink solutions emerge from the vacuum
expectation value of such an operator in  the Born approximation
and/or  $\la \rightarrow 0$ limit. The presented method is general
in the sense that applies also to the case of finite temperature
and to non--equilibrium; it also allows for the determination of
Green's functions in the presence of topological defects. We
discuss the classical kink solutions at $T\neq 0$ in the high
temperature limit. We conclude with some speculations on the
possible relevance of our method for the description of the defect
formation during symmetry--breaking phase transitions.\\

\vspace{3mm}
\end{abstract}

\draft
\vskip8mm

\section{Introduction}

\noi In recent years much attention has been devoted to the study
of topological defects and to the issue of their formation in the
course of symmetry--breaking phase transitions\cite{Kibble:1976sj,ESF}.
This
interest arises essentially for two reasons. On one hand,
topological defects are naturally inherent to many dynamical
systems ranging from cosmology to condensed matter physics, and
thus they provide a bridge between processes which manifest
themselves at very different energies and time scales
\cite{nature,Zurek:1996sj,volovik}. On the other hand, the study
of the formation of defects during phase transitions offers an
important way to understand better the underlying non--equilibrium
dynamics of quantum fields\cite{Boj,Zurek:2000ym}.

\vspace{3mm}

\noi The basic phenomenological picture of defect formation is
presently understood via the Kibble--Zurek
mechanism\cite{Kibble:1976sj,Zurek:1996sj}: In many systems, when
the phase transition associated with a spontaneous symmetry
breaking takes place, some regions of space may remain trapped in
the initial symmetric phase.  These regions are called {\em
topological defects}. In this way cosmic strings may have been
formed in the early universe; vortices in superfluids and
superconductors, and textures in some nematic liquid crystals are
also created in a similar fashion. A full understanding of the
process of defect formation in the context of Quantum Field Theory
(QFT) is however still far to be reached, although recently there
has been much progress in this
direction
\cite{Boj,Zurek:2000ym,calzetta,habib,laguna,rivers,rajantie}.
There are two fundamental points which make difficult such a task:
the emergence of the macroscopic (collective) behavior associated
with the defects in question resulting out of the underlying
microscopic (quantum) dynamics, and the intrinsic off--equilibrium
character of the processes  - realistic phase transitions - in
which the defects are created.

\vspace{3mm}

\noi In the present paper, we address the first question, i.e. we
show how a self--consistent description of topological defects can
be achieved in QFT in a way that is suitable for the study of
their formation in non--equilibrium processes - task which we plan
to tackle in our future  work. We are inspired by an approach to
the description of extended objects (solitons) as condensates of
particles in QFT which was developed in 70's by Umezawa, Semenoff
et al. \cite{UMZ1,UMZ2}. In their approach, known as the {\em
boson transformation method}, the usual order of argumentation is
reversed: solitons emerge naturally as macroscopic objects with
the full quantum origin. The classical soliton solutions are then
recovered in the $\hbar \rar 0 $ limit. Here we revisit this
method and reformulate it in the framework of the Closed-Time-Path
(CTP) formalism \cite{Blasone:1999aq}.  There are several
advantages in doing this. First of all, we obtain a closed
 functional form for the Heisenberg picture field operator in
the presence of defects. This is to be compared with the original
approach in which such an expression is obtained recursively via the
Yang--Feldman equation. The most important feature of the presented
formulation relies however on the fact that the CTP formalism allows
for a unified treatment of both the zero temperature and the finite
temperature cases. It is also suitable for the treatment of
non--equilibrium situations: in this case, as it is well known, the use
of CTP is essential \cite{CHu}. Last but not least, our formulation
allows to determine Green's functions in the presence of defects, which
can be useful for example for setting up self--consistently the Cauchy
data for a non--equilibrium evolution \cite{Boyanovsky:1998ka,Jizba:1998ig}.

\vspace{3mm}

\noi In this paper,
we consider a toy model system,
namely the (1+1)
dimensional $\la \psi^4$ theory,  both at zero and finite
temperature. We construct the Heisenberg field operator
 in the kink sector and recover the known (classical) kink
solutions in the classical limit as vacuum expectation values of
the field operator. We do this in
order to set the mathematical framework and explain techniques which
we are going to ``loosely replicate"  in our further work, where more
realistic systems will be considered \cite{toppress}.

\vspace{3mm}

\noi An interesting consequence of the present method are the
recurrence relations we arrive at in Section III,
which reveal an intimate connection between the
(non--exact) solvability of the ($1+1$) $\lambda \psi^{4}$ theory and
the solutions of Cauchy--Marley's functional equations.  In this paper
we do not consider quantum corrections to the classical soliton
solutions, although the method here presented allows for a
systematic study of this aspect (see comments in Section III), which
will be treated elsewhere.

\vspace{3mm}

\noi We then go one step further and show how to extend the
treatment for a system in thermal equilibrium saturated at some
temperature $T$. We show that in the high $T$ limit it is indeed
possible to get analytical solutions for the (inhomogeneous) order
parameter. A delicate limit $\hbar \rightarrow 0$ at finite $T$ is
discussed and compared with other existing treatments
\cite{vitman}. The results obtained are of general validity and
can be applied, for instance,  to more realistic higher
dimensional cases for the study of restoration of symmetry.

\vspace{3mm}

 \noi The plan of the paper is as follow:  In Section II  we
briefly review  the boson transformation method and the Haag
expansion which are the basic ingredients of our method.
As we aim to study topological defects in both equilibrium and
non--equilibrium media we formulate the whole approach in the CTP
formalism.

\vspace{3mm}

\noi In Section III we consider  the zero--temperature (1+1) $\la
\psi^4$ theory in  the Goldstone phase. We construct the {\em
classical} kink solutions  from the vacuum expectation value of
the Heisenberg picture field operator, in the Born
approximation and/or the $\lambda \rightarrow 0$ limit. These
limits are realized via residue calculations of certain functional
relations.

\vspace{3mm}

\noi In Section IV we extend our study to finite temperature. To
outline the basic strategy we consider the case of large $T$: this
allows us to obtain in a closed form the solution for the (thermal)
order parameter. The high temperature analysis reveals that there
cannot be any phase transition at high $T$ but, instead, the
system is plagued by spontaneous fluctuations exemplified by the
non--analytic behavior in the order parameter. We also show
explicitly that one cannot talk about the loopwise expansion as
being an expansion  in $\hbar$ at finite temperature. This
essentially distinguishes our approach from the  ones where tree
approximations at finite T are considered as equivalent to
classical limit.

\vspace{3mm}

\noi We conclude in Section V with some comments and speculations
on the possible relevance of the presented method for the description
of defect formation during phase transition. We also discuss the
r{\^o}le of the shift function (used for a construction of the
inhomogeneous coherent states) as a possible alternative tool for
the classification of topological  defects.

\vspace{3mm}

\noi In the Appendices are clarified some finer mathematical
manipulations needed in the main body of the paper.

\section{Description of solitons as inhomogeneous particle condensates }

\noi It has been recognized since long time that due to the breakdown
of von Neumann's uniqueness theorem \cite{Haag} the structure of
Quantum Field Theory is extremely rich, allowing for a host of
different (mutually unitarily inequivalent)  Hilbert spaces for  a
given dynamics of Heisenberg fields.  The choice of the  Hilbert
space is equivalent to fixing the boundary conditions for the
(operatorial) Heisenberg equations. It should be born in mind that a
particular choice of boundary conditions specifies the {\em
observational} particle (or quasi--particle) content of the theory.
In the perturbative Lehmann--Symanzik--Zimmermann (LSZ) approach, the
boundary conditions are injected via the asymptotic fields (most
commonly via the $in$--fields), and the functional relation between
them and the Heisenberg fields is then known as {\em Haag's map}.  One
practically realizes such a map expanding the Heisenberg fields into a
(functional) power series  of the asymptotic fields -
the {\em Haag's expansion}\footnote{The Haag expansion is often
defined as a power series of a normal ordered product of the
asymptotic fields\cite{garba}.  The connection between the definition
we use and the one in\cite{garba} can be established via the
operatorial Wick's theorem.}.

\vspace{3mm}

\noindent It is worth to stress that Haag's expansion is not an
operatorial relation, but holds true only in a weak sense, i.e.
when matrix elements are considered. As we shall see in Section II.B,
this is due to the fact that the Hilbert spaces for the Heisenberg
fields and asymptotic fields are mutually unitarily inequivalent.

\subsection{The ``boson transformation'' method}

\noi Starting from these basic observations, an approach to the
description of extended objects (solitons) as condensates of
particles in QFT was developed in the seventies by Umezawa et al.
\cite{UMZ1,UMZ2}.  In this approach solitons emerge naturally as
macroscopic objects (inhomogeneous condensate) with the full
quantum origin, provided one chooses the ``right" Hilbert space
for the asymptotic fields. The classical  soliton solutions are
then recovered in the Born approximation.

\vspace{3mm}

\noi Let us here briefly recall the main lines
of the boson transformation method.
To avoid unnecessary difficulties we will
ignore, temporary, the renormalization problem. Let us consider
the simple case of a dynamics involving one scalar field $\psi$
satisfying the equation of motion:

\bea \lab{bt1} \La(\pa) \, \psi(x)\, =\, {\mathcal{J}}[\psi](x) \, ,  \eea
where  $\La(\pa)$ is a differential operator, $x\equiv
(t,{\mathbf{x}})$ and ${\mathcal{J}}$ is   some functional  of the $\psi$ field,
describing the interaction. Let us now denote the asymptotic field
by $\varphi(x)$, satisfying  the equation
\bea \lab{bt2} \La(\pa) \, \varphi(x)\, = \, 0 \, .  \eea
Equation (\ref{bt1}) can be formally recast in the following
integral form (Yang--Feldman equation):
\bea \lab{bt3}
\psi(x)\, =\, \varphi(x)\, + \, \La^{-1}(\pa)\ast \,{\mathcal{J}}[\psi](x) \,
,  \eea
\noindent where  $\ast$  denotes convolution. The symbol $\La^{-1}(\pa)$
denotes formally  the   Green function for the $\varphi(x)$  field.  The
precise form of Green's function is specified by the boundary
conditions.  Eq.(\ref{bt3}) can  be  solved by iteration, thus giving an
expression for the Heisenberg  field $\psi(x)$ in terms of powers of the
$\varphi(x)$ field; this is the Haag expansion (or ``dynamical map'' in
the language of refs.\cite{UMZ1,UMZ2}), which might be formally written as
\bea  \lab{bt4} \psi(x)\, =\, F \lf[x; \varphi \ri] \, .  \eea
As already remarked, such an expression is valid only in a weak sense,
i.e. for the matrix elements only. This implies that
 Eq.(\ref{bt4}) is not
unique, since different sets of asymptotic fields (and the
corresponding Hilbert spaces) can be used in the construction. Let us
indeed consider a c--number function $f(x)$, satisfying the same free
equations of motion:
\bea \lab{bt5}
\La(\pa) \, f(x)\, = \, 0 \, ,
\eea
then the corresponding Yang--Feldman equation takes the form
\bea \lab{bt6}
\psi^f(x)\, =\,  \varphi\,  + \, f+ \, \La^{-1}(\pa)
\ast {\mathcal{J}}[\psi^f](x)\, .
\eea
\noindent The latter gives  rise to a {\em  different} Haag expansion
for a field  $\psi^f(x)$ still satisfying the Heisenberg equation
(\ref{bt1}):
\bea \lab{bt7} \psi^f(x)\, =\, F \lf[x; \varphi +f\ri]\, .  \eea
The difference between the two solutions $\psi$ and $\psi^f$ is only in
the boundary conditions. An important point is that the expansion
Eq.(\ref{bt7}) is obtained from that in Eq.(\ref{bt4}), by the
space--time dependent translation
\bea
\lab{bt7b} \varphi(x) \,\rar \,\varphi(x) \,+\,f(x)\, .
\eea

\noindent    Eqs.(\ref{bt7}) and (\ref{bt7b})   express   the essence of
the so called,  {\em boson transformation theorem} \cite{UMZ1}: the
dynamics embodied in Eq.(\ref{bt1}), contains an internal freedom,
represented by the possible choices of the function $f(x)$, satisfying
the  free field  equation
(\ref{bt5}).  Also observe  that the transformation
(\ref{bt7b}) is a canonical transformation since it leaves invariant the
canonical form of commutation relations.

\vspace{3mm}

\noindent   The  vacuum expectation value   of  Eq.(\ref{bt6}) gives
($|0\ran $ denotes the vacuum for the free field $\varphi$):
\bea \lab{bt8}
\phi^f(x)&\equiv & \,\lan 0| \psi^f(x) |0\ran \; = \;
f + \,\lan 0|\lf[ \La^{-1}(\pa) \ast {\mathcal{J}}
[\psi^f](x) \ri]|0\ran \, .  \eea

\noindent   Notice that the  order  parameter $\phi^f(x)$ is of
full quantum nature: the classical solution is obtained by means
of the classical or Born approximation, which consists in taking
$\,\lan 0|{\mathcal{J}}[\psi^f]|0\ran= {\mathcal{J}}[\phi^f]$,
i.e. neglecting all
contractions of the physical fields. In this limit,
$\phi^f_{cl}(x)\equiv \lim_{\hbar\rar 0}\phi^f(x) $ is the
solution of    the classical Euler--Lagrange equation:
\bea \lab{bt9}
\La(\pa) \, \phi^f_{cl}(x)\, =\, {\mathcal{J}}[\phi^f_{cl}](x) \, .
\eea

\noindent Beyond the classical level, in general,  the form of  this
equation changes. The Yang--Feldman equation (\ref{bt6}) describes
not  only the equations for the order parameter,
i.e. Eq.(\ref{bt9}), but also, at higher orders
in $\hbar$, the dynamics  of one or more quantum physical particles in
the potential generated  by the macroscopic object $\phi^f(x)$
\cite{UMZ1}. Typical examples of interest include for instance a
scattering of quasi--electrons on the Abrikosov vortices in type--II
superconductors or scattering of second sound waves (thermal phonons) on
the vortices in superfluid $^{4}$He.

\vspace{3mm}

\noindent
In refs.\cite{UMZ1,UMZ2}, it is shown that the class of solutions of
Eq.(\ref{bt5}) which leads to non--trivial (i.e. carrying a non--zero
topological charge) solutions of Eq.(\ref{bt9}), are those which have
some sort of singularity with respect to Fourier transform. These can be
either  {\em divergent singularities} or {\em topological
singularities}.  The first are associated to a divergence of $f(x)$ for
$|{\mathbf{x}}|=\infty$, at least  in some direction.  The second means that
$f(x)$ is not single--valued, i.e. it is path dependent.  In both cases,
the macroscopic  object   described by the order parameter, will carry a
non--zero topological charge.

\vspace{3mm}

\noi It is also interesting to consider the boson
transformation at level of states \cite{Mercaldo:1981ip}. For this
purpose  let us write the generator of the field shift (\ref{bt7b}), as
\begin{eqnarray}
\non &&\varphi^f(x) \, =\, e^{-i {\cali D}}\,\varphi(x)\,
e^{i{\cali D}} \,=\,  \varphi(x)\, + \, f(x) \\ \lab{gencs} &&{\cali
D}\, =\, - \int \,f(y) \stackrel{\lrar}{\pa_{\mu}}\varphi(y)\,
d\sigma^{\mu} \, .
\end{eqnarray}
\noi with $d\sigma^{\mu} = n^{\mu} d\sigma$ where $\sigma$ is a flat
space--like surface  and $n^{\mu}$ its normal, both
$x$ and $y$ belong to $\sigma$.
The action of ${\cali D}$ on the vacuum defines a coherent state $|f\ran$:
\begin{equation}
\lan f| \varphi(x) |f \ran \, = \, f(x)\, , \qquad  \qquad | f\ran
\, = \, e^{i{\cali D}} \, |0\ran \, . \label{cs1}
\end{equation}
The classical soliton solution is then obtained by taking the
coherent--state expectation value of the Heisenberg field $\psi$
in the $\hbar \rightarrow 0$ limit, i.e.
\begin{equation}
\lim_{\hbar \rightarrow 0}\lan f| \psi(x) |f \ran \,  =
\lim_{\hbar \rightarrow 0}\, \lan 0| \psi^f(x) |0 \ran \,= \,
\phi^f_{cl}(x) \, . \label{coh3}
\end{equation}
\noi Although ${\cali D}$ in (\ref{gencs}) is a
precise analog of the Weyl operator of the quantum mechanics defining
the canonical coherent states\cite{JK1},
the states $|f\rangle$ defined by (\ref{cs1}) are not the usual
coherent states. Indeed, due to the fact that $f(x)$ is
not Fourier transformable, $|f\rangle $ are not eigenstates of the
annihilation operator $a({\bf k})$ albeit they still do saturate
the Heisenberg uncertainty relations.
 A discussion of  the coherent states
corresponding to the soliton solutions can be found, for example,
in refs.\cite{Taylor:1978bq,KCa1}. Connection of QFT coherent
states with Haag's theorem is discussed in\cite{JK2}.

\subsection{The Haag expansion in the Closed--Time Path formalism}
%
\noindent Whilst the Yang--Feldman equations  are quite involved and do
not usually  allow to proceed beyond few iterations, the
alternative  Heisenberg  equations seems to   be  more versatile.  This
may be  seen by considering the bare Heisenberg
equations of the motion:
\bea\lab{cpt0a} &&\dot{\psi}_{B}(x)    =  i[H,\psi_{B}(x)] \\
\lab{cpt0b} &&\dot{\Pi}_{B}(x) = i[H, \Pi_{B}(x)]\, , \eea
\noindent where $\Pi_{B}$ is the bare momentum conjugate to the
bare field $\psi_{B}$ and $H$ is the full  Hamiltonian   in the
Heisenberg picture ($H = \int   d^{3}x \,  {\cal H}(x)$).   The
formal solution  of (\ref{cpt0a})--(\ref{cpt0b}) is  well known.
Assuming that the Heisenberg and interaction pictures coincide at
some time $t_{i}$, we have\cite{Coll}:
\bea\lab{ks0a} \psi_{B}(x)&=& Z_{\psi}^{1/2}\La^{-1}(t)\; \psi_{in}(x)\;
\La(t) \\ \lab{ks0b} {\Pi}_{B}(x)&=& Z_{\Pi}^{1/2}\La^{-1}(t)\;
\Pi_{in}(x)\;\Lambda(t) \\ \lab{ks0c}
\Lambda(t)&=&e^{i(t-t_{i})H_{in}^{0}}\; e^{-i(t-t_{i})H} \\ \lab{ks1}
\Lambda(t_{2})\; \Lambda^{-1}(t_{1}) &=& U(t_{2};t_{1}) \, =\,
T\left[\exp(-i\int_{t_{1}}^{t_{2}}\; d^{4}x \, {\cal
H}_{in}^{I}(x))\right]\, .   \eea
\noindent $T$ is the usual time ordering  and $Z_{\psi}$,
$Z_{\Pi}$ are the wave--function renormalizations (usually $\Pi
\propto \dot{\psi}$, and so $Z_{\psi} = Z_{\Pi}$). Due to the fact
that both $\psi_{B}$ and $\psi_{in}$ satisfy the canonical
equal--time commutation relations, the solution $\psi_{B}$ in
(\ref{ks0a}) must be  understood in  a weak sense, i.e.  valid
for each  matrix element separately.  If   not,  we would obtain
 the  canonical commutator between $\psi_{B}$  and $\Pi_{B}$
being equal to $iZ_{\psi}\de^{3}({\bf  x} -{\bf y})$ and thus
canonical quantization would require that $Z_{\psi}=1$. On  the
other hand,  non--perturbative considerations (e.g. the
K\"allen-Lehmann  representation) require $Z_{\psi}<1$.  The
solution of this problem is well known\cite{Haag,GJ,IZ}: the
Hilbert spaces for $\psi_{B}$ and $\psi_{in}$ are   different
(unitarily non--equivalent), and the wave function
renormalizations $Z_{\psi}$ and/or $Z_{\Pi}$ are then
``indicators''  of how much  the unitarity is violated. This
conclusion is usually referred as Haag's theorem\cite{Haag,Ster}.

\vspace{3mm}

\noindent Let us also   mention  that the  interaction    picture
evolution operator $U(t_{2},t_{1})$ alternatively reads \cite{N}
\bea\lab{EO1} U(t_{2},t_{1})
&&=T^{*}\left[\exp(i\int^{t_{2}}_{t_{1}}\;d^{4}x \; {\cal
L}_{in}^{I}(x))\right], \eea
\noindent    where  the    symbol    $T^{*}$  is   called  $T^{*}$ (or
covariant) product\footnote{The $T^{*}$ product is  defined  in such a
way that for fields in the interaction picture it is  simply the $T$
product with all the derivatives  pulled out of the  $T$-ordering
symbol.  Evidently, for free fields without derivatives:
$ T^{*}[\psi_{in}(x_{1})\ldots \psi_{in}(x_{n})  ]   =
T[\psi_{in}(x_{1})\ldots \psi_{in}(x_{n})]. $}
\cite{IZ,N,RJ} and  ${\cal  L}_{in}^{I}$ is the interacting part  of the
density of   the Lagrangian in  the interaction picture.
 Eq.(\ref{EO1}) is  valid even  when the derivatives of
fields  are present in   ${\cal L}^{I}$ (and  thus ${\cal H}^{I} \not=
-{\cal L}^{I}$).

\vspace{3mm}

\noindent Eq.(\ref{ks0a}) can be recast into a more useful
form, namely
\bea\lab{cptdm} \psi_{B}(x)  &=&  Z_{\psi}^{1/2}\,U(t_{i};t)\,
\psi_{in}(x)\, U^{-1}(t_{i};t)
\,=\,Z_{\psi}^{1/2}\,U(t_{i};t_{f})U(t_{f};t)\,\psi_{in}(x) \,
U(t;t_{i}) \non  \\ \non  \\ &=& Z_{\psi}^{1/2}  \,T_{C}  \lf[
\psi_{in}(x) \exp\lf( -i \int_C d^4x\,  {\cal H}_{in}^{I}(x) \ri)\ri]
\,=\,Z_{\psi}^{1/2} \,T_{C}^{*} \lf[ \psi_{in}(x) \exp\lf(i \int_C
d^4x\, {\cal L}_{in}^{I}(x) \ri)\ri]\,.  \eea
\noindent  Here  $C$ denotes    a closed--time  (Schwinger) contour,
running from  $t_{i}$ to a later  time $t_{f};\, t \leq t_{f}$ and back
again (see Fig.1).  Similarly, $T_{C}$ denotes the corresponding
time--path ordering  symbol (analogously for   the $T^{*}_{C}$
ordering).  In  the limit $t_{i} \rar -\infty$, we  get that $\psi_{in}$
turns out  to be the usual   in--(or asymptotic) field.  As the time
$t_{f}$ is  by construction arbitrary, it is useful, from a technical
point of view, to set $t_{f}= +\infty$. Eq.(\ref{cptdm}) may be viewed
as the Haag  expansion of the Heisenberg field $\psi_{B}$.

\begin{figure}[t]
\vspace{-1cm} \centerline{\epsfysize=2.5truein\epsfbox{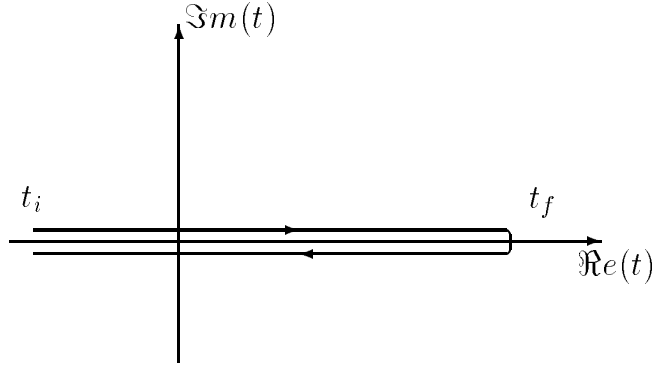}}
\vspace{-0.6cm}
\caption{The Closed Time Path}

\vspace{0.2cm} \hrule
\end{figure}

\vspace{3mm}

\noindent Generalization of Eq.(\ref{cptdm}) to the case of more fields
is straightforward. So, for instance, for the time ordered product of
$n$ Heisenberg fields we may write
\bea\lab{cptdm1} T[\psi_{B}(x_{1}) \ldots \psi_{B}(x_{n})] =
Z_{\psi}^{n/2}T^{*}_{C}\lf[ \psi_{in}(x_{1}) \ldots \psi_{in}(x_{n}) \,
\exp\lf(i\int_{C}d^{4}x \, {\cal L}^{I}_{in}(x)\ri)\ri]\, .  \eea
\noindent When we consider {\em vacuum expectation value} of Eq.(\ref{cptdm1}),
we have at our disposal two equivalent representations for the $T=0$
Green's functions, namely
\bea\lab{cptdm2a}
\langle 0| T(\psi_{B}(x_{1})\ldots \psi_{B}(x_{n}) )
|0 \rangle &=& Z_{\psi}^{n/2}\langle 0 | T^{*}_{C}\lf[
\psi_{in}(x_{1}) \ldots \psi_{in}(x_{2}) \, \exp\lf( i\int_{C}
d^{4}x \, {\cal L}^{I}_{in}(x)\ri)\ri] |0 \rangle
\\[4mm] \lab{cptdm2b}
&=&  Z_{\psi}^{n/2} \frac{\langle 0 |T^{*}\lf[ \psi_{in}(x_{1})\ldots
\psi_{in}(x_{n})\, \exp\lf( i \int_{-\infty}^{\infty} d^{4}x \,
{\cal L}^{I}_{in}(x)\ri)\ri] |0 \rangle  }{\langle 0| T^{*}\lf[
\exp\lf( i\int_{-\infty}^{\infty} d^{4}x \,  {\cal
L}^{I}_{in}(x)\ri)\ri] |0 \rangle }\, .  \eea
\noindent Eq.(\ref{cptdm2b}) is the well known Gell-Mann--Low formula
for Green's functions. Note that the latter is
true {\em only} for the vacuum expectation values, due to the stability
of the vacuum  $| 0 \rangle $:
\begin{displaymath}
T^{*} \left[\exp\lf( i \int_{-\infty}^{\infty} d^{4}x \, {\cal
L}^{I}_{in}(x) \ri)\ri]|0 \rangle = \alpha \, |0 \rangle \, ,
\end{displaymath}
with $\alpha $ being an eigenvalue (basically a phase factor)
of the interaction--picture evolution operator which corresponds to $| 0
\rangle$.  In more general situations where expectation values are taken
with respect to a state
$|\psi \rangle \not= |0 \rangle $,
Green's function cannot be recast in the form  (\ref{cptdm2b}) and
the $T^{*}_{C}$ prescription is obligatory. So for mixed states,
$|\psi \rangle \rightarrow \rho $,  $\langle
\psi | \ldots |  \psi \rangle \rightarrow Tr\lf( \rho \ldots \ri) =
\langle \ldots \rangle  $ and
\begin{equation}\lab{CTP1}
\langle T(P[\psi_{B}]) \rangle = \left\langle T^{*}_{C} \lf[
P_{r}[\psi_{in}] \, \exp\lf( i\int_{C} d^{4}x \, {\cal
L}^{I}_{in}(x) \ri) \ri]  \right\rangle \, ,
\end{equation}
where $P_{r}[\ldots]$ is an arbitrary (generally composite)
polynomial in $\psi_{B}$, and the subscript $r$ suggests that the
corresponding renormalization factors are included.  Eq.(\ref{CTP1})
will be of a fundamental importance in our following considerations.

\vspace{3mm}

\noindent An important special case is that of  a system in
thermodynamical equilibrium.  Then  the statistical properties of the
system are described by the canonical density matrix (for simplicity we
 omit from our consideration grand--canonical ensembles).
 As $\rho \propto e^{-\beta H}$, the
density matrix is basically a generator of the (imaginary) time
translations. Using Eq.(\ref{ks0c}) we may then  write
\begin{displaymath}
e^{-\beta H} = e^{-\beta H^{0}_{in}}\, U(t_{i}-i\beta ,  t_{i})\, .
\end{displaymath}
{}From this it is evident that one may substitute the full density
matrix with the density matrix for  the corresponding free system
provided one adds to the path $C$ a vertical part running from
$t_{i}-i\beta$ to $t_{i}$ (see Fig.2).  Advantage of this rather
formal step is that the free density matrix is Gaussian and
correspondingly $\langle T_{C}[\psi_{in}(x_{1})\ldots
\psi_{in}(x_{n})] \rangle $ is $0$ for $n$ odd and a symmetrized
product of the (free) two--point Green's function if $n$ is even.
This is nothing but the thermal (or thermodynamical) Wick's
theorem - basis for a perturbation calculus in QFT at finite
temperature. We shall elaborate more on this point in Section IV.

\begin{figure}[t]
\vspace{-1cm} \centerline{\epsfysize=2.5truein\epsfbox{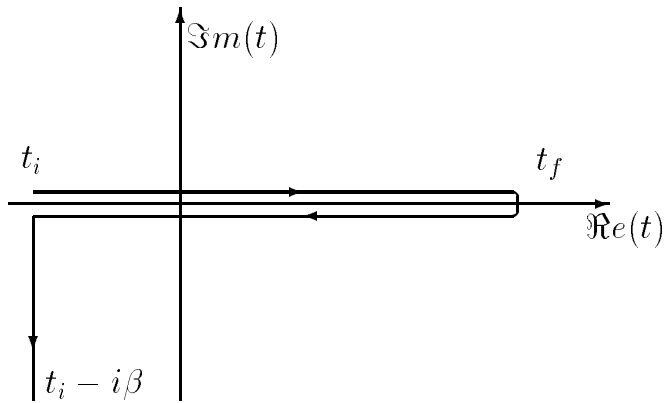}}
\vspace{-0.6cm}
\caption{The Thermal Path}

\vspace{0.2cm} \hrule
\end{figure}

\subsection{Heisenberg operator in the presence of defects}

\noi Having obtained a compact expression for the Heisenberg field in
terms of the interacting fields, we can select the initial--time
conditions corresponding to a particular physical situation. In other words
(see Section II.A), we select the set
of asymptotic fields describing the particle content of the theory in
question. For instance, in the case of a type--II superconductor we must
choose either quasi--electrons, if we are above the critical temperature
(normal phase), or bogolons if we are bellow the critical temperature
(superconducting phase)\cite{RDMa}. It is well known that the new vacuum
state in the superconducting phase is an (homogeneous) condensate of Cooper's
pairs (BCS state).

\vspace{3mm}

\noindent Similarly, for paramagnetic--ferromagnetic phase
transition in the Heisenberg model we choose particles (atoms)
with spin $\frac{1}{2}$ above the Curie temperature (paramagnetic
phase) whilst we choose magnons if we are below the Curie
temperature (ferromagnetic phase). The vacuum in the ferromagnetic
phase is then a state where all particles have spins aligned in
the same direction - an (homogeneous) spin (or Bloch) coherent
state.

\vspace{3mm}

\noindent Let us also mention the important case of the QCD phase
transition. Here, in the confined phase (i.e. at energy scale $\ll
\, \Lambda_{QCD}$) one usually chooses as asymptotic states
hadrons and mesons (i.e. quark bound states). In the deconfined
phase, i.e. at temperatures of order $\sim 0.2 \mbox{GeV} \sim
10^{12}K$ where a quark gluon plasma starts to form and the chiral
symmetry is restored, one could, in principle, directly work with
quarks and gluons as relevant asymptotic fields. In this case, the
structure of the confined phase vacuum is still far from being
understood. The latter is sometimes identified with the
(homogeneous) chiral condensate or with ``color'' BCS state, etc.

\vspace{3mm}

\noi The choice of the initial--time data is therefore an important
theoretical probe for the description of phase transitions and
their underlying microscopic mechanism. If a phase transition is
sufficiently rapid it may happen that  stable topological defects are
created; examples include vortices in rotating superfluid $^{4}$He below
the $\lambda$ point, a rich variety of defects in $^{3}$He, quantum
magnetic flux tubes (Abrikosov vortices)  in type--II superconductors,
disclination lines and other defects in liquid nematic crystals,
droplets of an unbroken phase in quark gluon plasma, etc. As we have
already mentioned in Section II.A, such topological defects can be
generated with a
special choice of the initial--time data by means of the boson
transformation  with a Fourier non--transformable shift function $f$.
Loosely speaking, the
emergence of topological defects may be seen in the corresponding vacuum
state for the asymptotic fields which should be
a ``suitably chosen"  inhomogeneous coherent state.

\vspace{3mm}

\noi We can summarize our strategy as follows (see Fig.3):
For a given dynamics, the CTP
formulation gives a closed (functional) expression for the Heisenberg field
operator(s) $\psi$ in terms of the asymptotic (physical) fields. We then use the
boson transformation to introduce the shift function $f$
controlling the choice of the
Hilbert space.

\begin{figure}[t]
\vspace{-1cm} \centerline{\epsfysize=3.0truein\epsfbox{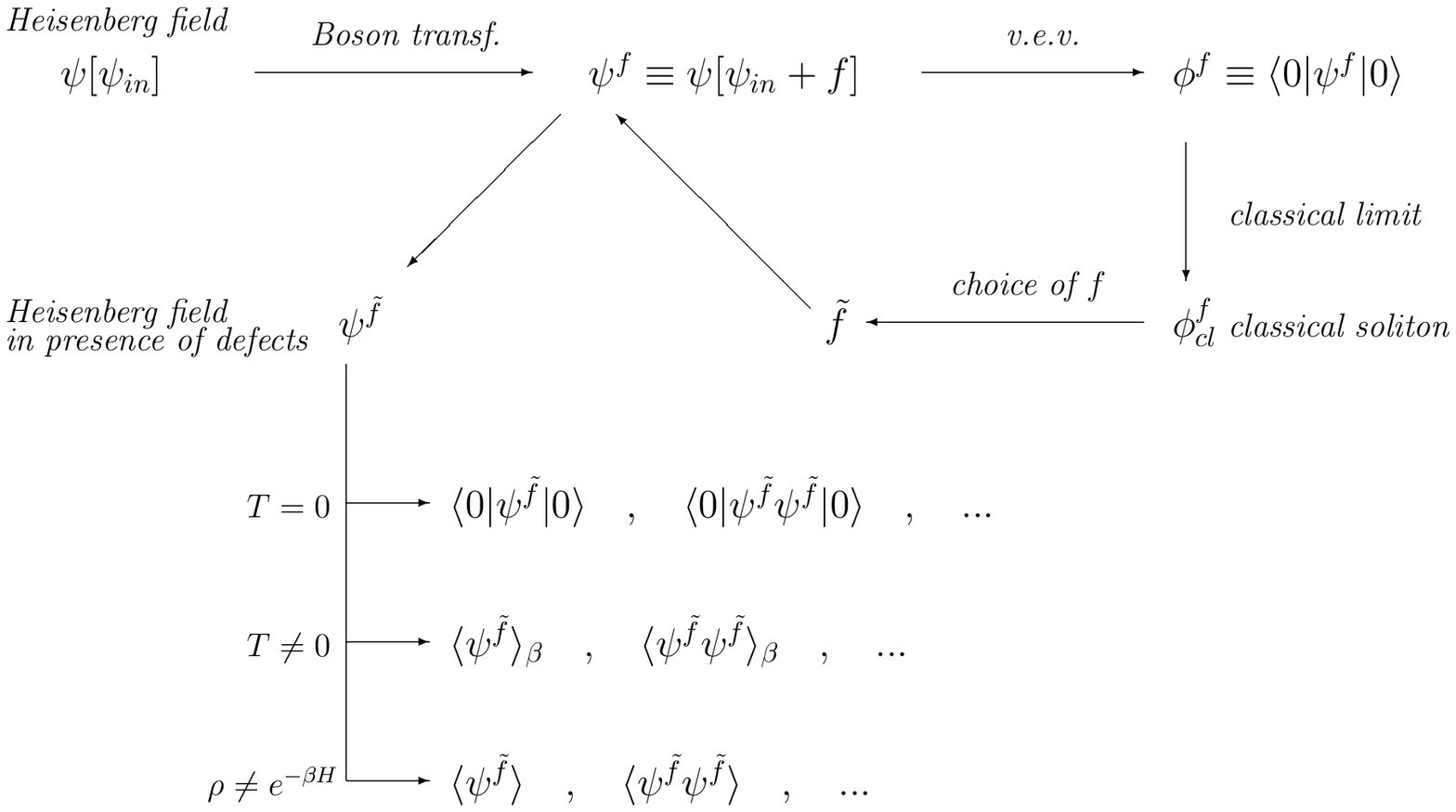}}
\vspace{0.6cm}
\caption{Iterative procedure for computations}

\vspace{0.2cm} \hrule
\end{figure}

\vspace{3mm}

\noi The next step is to consider the order parameter, i.e.
$\phi^f\equiv \lan 0 |\psi^f|0\ran$. By taking the classical limit
$\phi_{cl}^f\equiv \lim_{\hbar\rar 0} \phi^f $, we determine the
form of the shift function, say ${\ti f}$, corresponding to a
particular (classical) soliton solution. This shift function is
then used to obtain the Heisenberg field operator in the chosen
soliton sector: $\psi^{\ti f}$. At this point there are various
possibilities:

\vspace{3mm}

\noi - calculate quantum corrections to the order parameter, by taking
higher orders in the $\hbar$ expansion of $\lan 0|\psi^{\ti f}|0\ran$;

\vspace{2mm}

\noi - study finite temperature effects on the order parameter,
by considering
$\lan \psi^{\ti f} \ran_{\bt}$,
where $\lan ... \ran_{\bt}$ stands for thermal average;

\vspace{2mm}

\noi - calculate Green's functions in the presence of defects, both at zero and
finite temperature, as $\lan T[\psi^{\ti f} \psi^{\ti f} ]\ran$, etc.

\vspace{3mm}

\noi A more ambitious task is the study of the non--equilibrium
properties of QFT systems
containing defects, e.g. consider phase transitions.
This will be the object of our future work.

\section{Kinks in two--dimensional $\la \psi^4$ theory at $T=0$}

\noindent In this Section we apply the formal considerations developed
above to a specific model, namely the two--dimensional $\la \psi^4$
theory at zero temperature.
%
It is well known that Goldstone's sector of the $\lambda\,
\psi^{4}$ theory in $(1+1)$ dimensions possesses at a classical
level a certain class of analytical extended solutions,
namely the kink (antikink)
solutions \cite{Rajaraman:1982is}. In the following, we
derive the classical kink solutions
at $T=0$ and construct the Heisenberg operators in presence of kinks.
As a further non--trivial step, we extend in the next
Section this analysis to finite temperature.

\subsection{The Heisenberg operator in the presence of a kink}
%
\noindent  Let us consider  the case of  a $1+1$ dimensional scalar
system of the hermitian field $\psi$  with quartic interaction term.  We
have the following bare Lagrangian  (throughout we adopt the Minkowski
metric with signature $(+\, -)$):
\begin{eqnarray} {\cal  L}\,   &=&\,   \frac{1}{2}\,(\pa_\mu   \psi_{B})^2
- \frac{1}{2}\, \mu_{B}^2 \psi_{B}^2 - \frac{\la_{B}}{4}\, \psi_{B}^4
\nonumber \\ [2mm]
&=&  \frac{1}{2}\, (\pa_\mu   \psi)^2 - \frac{1}{2}\, \mu^2
\psi^2 - \frac{\la}{4} \,\psi^4 \; + \; {\cal L}_{ct} \, ,
\label{lagr1}
\end{eqnarray}
\noindent where ${\cal L}_{ct}$ is the counterterm Lagrangian:
\begin{displaymath}
{\cal L}_{ct} = \frac{1}{2}(Z_{\psi} - 1)\, (\pa_{\mu} \psi)^{2} -
\frac{1}{2}\, \delta \mu^{2} \psi^{2} - \frac{\lambda}{4}\, (Z_{\lambda}
- 1)\psi^{4}\, .
\end{displaymath}
\noindent Here we have introduced in the usual way a renormalized field,
mass and coupling as: $\psi_{B} = Z^{1/2}_{\psi}\psi$,
$Z_{\psi}\mu_{B}^{2} = \mu^{2} + \delta \mu^{2}$ and $\lambda_{B} =
Z_{\psi}^{-2}Z_{\lambda} \lambda$. The corresponding renormalized
Heisenberg equation of the motion for the field operator $\psi$ is (see
e.g. \cite{Coll})
\begin{equation}
(\pa^{2} + \mu^{2})\psi(x) = - \la [\psi^{3}(x)]\, .
\label{pj34}
\end{equation}
\noindent The squared bracket on the RHS of (\ref{pj34}) denotes a
{\em renormalized} composite operator (or Zimmerman normal
ordering) corresponding to an unrenormalized operator $\psi^{3}$.
%

\vspace{3mm}

\noindent In the case when both $\mu^{2} < 0$ and  $\mu^{2}_{0} <
0$ (i.e. in the Goldstone phase)  it is well known that
(\ref{lagr1}) admits at a classical level kink
solutions\cite{Rajaraman:1982is,garba,Coleman}.  We shall omit from
the next discussion the Wigner phase (i.e. when  $\mu^{2}_{0}> 0$)
since it does not enjoy the kink solutions.  Let us define the
following renormalized quantities

\begin{equation}\psi(x) =   v + \rho(x) \qquad
{\mbox{with}} \quad v = \sqrt{-\frac{\mu^{2}}{\lambda}} \quad ,\quad
-2\mu^2 =  m^2 \quad,\quad  g=\sqrt{2 \la}\, .
\label{pj39}
\end{equation}
\noindent The parameter $v$ represents a tree--level expectation
value of $\psi$ (i.e. the true classical minimum of the potential
in (\ref{lagr1})). As a consequence of the parameterization
(\ref{pj39})  we obtain
\begin{equation} {\cal  L}\,   =\,  \frac{1}{2}\,(\pa_\mu   \rho)^2    -
\frac{1}{2}\, m^2 \rho^2 - \frac{g^2}{8} \,\rho^4 - \frac{1}{2}\, mg
\rho^3 + \frac{m^4}{8 g^{2}} \; + \; {\cal L}_{ct}\, ,
\label{pj38}
\end{equation}
\noindent where
\begin{eqnarray*}
{\cal L}_{ct} &=& \frac{1}{2}\, (Z_{\psi} - 1)(\pa_{\mu}\rho)^{2} -
\frac{1}{4}\, \left( (Z_{\lambda} - 1)\frac{3}{2}\, m^{2} - \delta m^{2}
\right)\, \rho^{2} - \frac{1}{2} \, g (Z_{\lambda} - 1)m \rho^{3} -
\frac{g^{2}}{8}\,(Z_{\lambda} -1) \rho^{4}\nonumber \\ &-& \frac{1}{2}
\, \frac{m}{g}\, \lf( \delta m^{2} - (Z_{\lambda} - 1)m^{2} \,\right)
\rho + \frac{1}{4}\, \frac{m^{2}}{g^{2}}\, \lf( \delta m^{2} -
\frac{1}{2}\, (Z_{\lambda} - 1 )m^{2} \right) \, .
\end{eqnarray*}

\noindent Here we have used that $\delta m^{2} = -2 \delta
\mu^{2}$. The form (\ref{pj38}) of the Lagrangian is particularly
important in doing perturbative theory. We may recall that both
$m$ and $g$ in ${\cal L}$ are clearly (by construction) arbitrary
parameters and no physical observable should depend on their
choice.
But, in practice, when one deals with perturbative calculations,
choice of one parameterization over another can save some labor
and possibly cultivate on the intermediate stage a better physical
intuition.  Here, and in the work to follow, we explicitly choose
the zero--momentum mass renormalization prescription: i.e.
$\Sigma_{R}[p^{2}=0] = -m$; with $\Sigma_{R}$ being the
renormalized self--energy. Similarly, we chose  the zero--momenta
coupling renormalization  condition: $\Gamma^{(4)}_{R}[0,0,0,0] =
-\lambda$; with $\Gamma^{(4)}_{R}$ being the renormalized
$4$--point proper vertex function.  Both $m$ and $\lambda$ are
chosen to be the tree level mass and coupling,  respectively. The
virtue of this prescription is that we keep from the very
beginning the  interpretation of $v$ as the tree--level vacuum
expectation
value of $\psi$. In addition, the zero--momentum renormalization
prescription can be easily recast into conditions on the
renormalized effective potential which is the most natural tool
for a treatment of phase transitions\cite{IZ,Peskin} (this is
particularly useful when comparing with existing results which are
predominantly computed in the effective potential framework). The
physical, but mathematically more involved on--shell
renormalization may be obtained via a finite renormalization
procedure\cite{IZ}. Let us also remark that in 2$D$ case the
$\lambda \, \psi^{4}$ theory is super--renormalizable and so both
$Z_{\lambda}$ and $Z_{\psi}$ are finite to each order in the
perturbative expansion, while the divergence present in $\delta
m^{2}$ is solely due to the tadpole diagram which is here only
logarithmically divergent.

\vspace{3mm}

\noindent The
renormalized Heisenberg equation for the field $\rho$ is:
\bea (\pa^{2} + m^{2})\rho(x) &=&  -\frac{3}{2} m g [\rho^{2}(x)] -
\frac{1}{2} g^2 [\rho^{3}(x)] \, .  \eea
\noindent Note that if $\mu^{2} <0$ then $m^{2}>0$.  The asymptotic
($t\rar -\infty$) field can be then identified with the free massive
field  satisfying
\bea (\pa^{2} + m^{2})\rho_{in}(x) &=& 0 \, ,  \eea
\noindent with $m$ being the aforementioned tree level renormalized
mass.  Setting in Eq.(\ref{ks0c})
$H_{in}^{0}=\frac{1}{2}{\dot{\rho}}^{2}_{in}+
\frac{1}{2}{\rho'}^{2}_{in}-\frac{1}{2}m^2 \rho^{2}_{in}$ then the
interaction Lagrangian ${\cal{L}}^{I}$  (entering Eq.(\ref{EO1})) is
\begin{eqnarray*}
{\cal{L}}^{I}[\rho] = {\cal L}_{ct}[\rho]\; - \frac{1}{8}\, g^{2}
\rho^{4} - \frac{1}{2}\, m g \rho^{3} + \frac{m^{4}}{8g^{2}}\, ,
\end{eqnarray*}
\noindent and the dynamical map for the field $\psi$ has the form
\bea &&\psi(x) = v + T_{C}^{*}  \lf[ \rho_{in}(x) \exp\lf\{i
\int_C d^2y\, {\cal{L}}^{I}[\rho_{in}]\ri\}\ri]\, .  \eea

\noindent Note that the constant factor in ${\cal L}^{I}$
automatically cancels during the contour integration\footnote{This
fact, among others, shows that  in the CTP formalism the UV
divergence in the energy density of the vacuum is automatically
cancelled.}.  Let us now consider the boson transformation:
\bea \lab{cpt3a} && \rho_{in}(x) \, \rar \, \rho_{in}(x) + f(x) \\ \non
&& \\ \lab{cpt3b} &&(\pa^2 + m^2) f(x) \, =\,0 \, .  \eea

\noindent As a result we get the new Haag expansion for the field
$\psi^f(x)$
\begin{eqnarray} \lab{psif1}
 \psi^f(x) & =& \, v\, +\, \,T_{C}^{*}  \lf[ (\rho_{in}(x)\, +\, f(x))
\exp\lf\{i \int_C d^2y\, {\cal{L}}^{I}[\rho_{in}(y) +
f(z)]\ri\} \ri]  \nonumber\\ [2mm]
  & =& \,v\, + \lf[\frac{\de}{i\de J(x)}\,
+\, f(x) \ri]\,\exp\lf\{i \int_C d^2y\,
{\cal{L}}^{I}\lf[ \frac{\de}{i\de J(y)} + f(y)\ri]\ri\} \,
\lf.T_C\lf(\exp i \int_C d^2y\, J(y)\rho_{in}(y)\ri)\ri|_{J=0} \, ,
\end{eqnarray}
where we have introduced the $c$--number source $J$ in order
to perform some formal
manipulations.  By use of the (operatorial) Wick's theorem\cite{IZ}, we
get
\bea  \psi^f(x) &=&  v\, +\, \lf[\frac{\de}{i\de J(x)}\, +\, f(x) \ri]\,
 \exp\lf\{i \int_C d^2y\,{\cal{L}}^{I} \lf
 [\frac{\de}{i\de J(y)} + f(y) \ri]\ri\} \non \\[2mm]
  && \times \, :\exp i
 \int_C d^2y\, J(y)\rho_{in}(y): \,\lf.  \exp \lf[-\frac{1}{2}\int_C
 d^2y d^2z\,J(y)\De_C(y;z)J(z)\ri] \ri|_{J=0}\, , \lab{cpt72} \eea
\noindent where $\De_C (x;y)=\lan 0|T_C(\rho_{in}(x)\rho_{in}(y))|
0\ran$ and $|0\ran$ is the vacuum for the $\rho_{in}$ field.  Once
the function $f$ is ``properly'' chosen (see below),
Eq.(\ref{cpt72}) represents  a convenient representation for the
Heisenberg operator  in presence of defects which can be used for
further analysis\cite{Blasone:1999aq,toppress}.

\vspace{3mm}

\noindent In order to determine the function $f$ leading to kink
solutions,  let us now consider the vacuum expectation value  of the
Heisenberg field  $\psi^{f}$. Here the normal ordered term drops, and we
get ($\lan \ldots \ran\, \equiv\lan 0|\ldots| 0\ran $)
\bea  \lan \psi^f(x) \ran &=&  v\, +\,   \lf[\frac{\de}{i\de J(x)}\, +\,
f(x)\ri] \,\exp\lf\{i \int_C d^2y\, {\cal{L}}^{I}\lf
[\frac{\de}{i\de J(y)} + f(y) \ri]\ri\} \,
\nonumber \\[2mm]  &&\quad \times \;
\lf. \,\exp \lf[-\frac{1}{2} \int_C d^2y d^2z\, J(y)\De_C(y;z)J(z)\ri]
\ri|_{J=0}\, .  \lab{cpt73} \eea

\noindent By use of the relation\cite{Coleman}
\bea  F\lf[ \frac{\de}{i\de J} \ri] G[J] = \lf.  G\lf[ \frac{\de}{i\de
K} \ri] F[K] e^{i \int K J}\ri|_{K=0}\, , \eea
\noindent we obtain
\bea \non \lan \psi^f(x) \ran &=&  v\,+\,   \exp\lf[-\frac{1}{2}
\int_C d^2y d^2z\,\De_C(y;z) \frac{\de}{i\de K(y)}\frac{\de}{i\de K(z)}
\ri] \,\qquad  \\[2mm]
&&\times  \;\lf(K(x)\, +\, f(x) \ri)
\lf. \exp\lf\{i \int_C d^2y\,\lf[ {\cal{L}}^{I}[K(y) + f(y)]
+  K(y) J(y)\ri]\ri\} \ri|_{K=J=0}\, .
\eea

\noindent We now perform a change of variables $K(x) \rar K(x)\, +\,
f(x) $ and set to zero the term with $J$ (there are no derivatives with
respect to it). As a result we obtain
\begin{equation}
\lan \psi^f(x) \ran =  v\,+\, \lf. \exp \lf[-\frac{1}{2} \int_C d^2y
d^2z\,\De_C(y;z) \frac{\de}{i\de K(y)}\frac{\de}{i\de K(z)} \ri]
K(x)\, B[K] \ri|_{K=f}\, ,
\label{cpt10}
\end{equation}
\noindent with
\bea
B[K] \equiv \exp\lf\{i \int_C
d^2y\,{\cal{L}}^{I}[K(y)]\ri\}\, .
\eea

\pagebreak

\noindent We can thus express Eq.(\ref{cpt10}) as a sum of three terms
\footnote{We use the  identity
\bea \non
 \exp\lf[\frac{1}{2}\sum_{ij} \De_{ij} \pa_{x_i}
\pa_{x_j}\ri] \, x_k
\, B(x_l) =
\lf(x_k + \sum_{j} \De_{kj}\pa_{x_j}\ri) B(x_l + \sum_{j} \De_{lj}\pa_{x_j})
\ident \,,  \eea
where $x_k \rar K(x) $ and $ B(x) \rar B[K] $.}
\begin{eqnarray} \lab{3parts}
\lan \psi^f(x)\ran &=&  v \, +\,\lf.   C[K](x)\ri|_{K=f} \, +\, \lf.
  D[K](x)\ri|_{K=f}  \\ && \non  \\
C[K](x) &= & \int_{C} d^{2}y \De_{C}(x;y) \frac{\de}{\de K(y)} \,
\exp{\lf[\frac{1}{2} \int_{C} d^{2}z d^{2}y \De_{C}(z;y)
\frac{\de^{2}}{\de K(z) \de K(y)}\ri]}\,B[K] \, ,
   \\ &&\non  \\
 D[K](x) &=& K(x)\, \exp\lf\{\frac{1}{2} \int_{C}d^{2}x d^{2}y
\De_{C}(x;y) \frac{\de^{2}}{\de K(x) \de K(y)}\ri\} \, B[K]
\,    .
\end{eqnarray}

\noindent Let us observe that
$C[K=0](x)$ is independent on $x$ since the  inhomogeneity in the
order parameter is controlled by the $x$ dependence in $K(x)$  (or
precisely in $f(x)$). As a result we see that $C[0]$ describes
quantum  corrections to the tree level {\em homogeneous} density
of condensation $v$  (note $D[K=0] = 0$).  Introducing the
notation:
\bea {\cal C}[K](x) \equiv C[K](x) - C[0] \, , \;\;\;\; {\tilde{v}}
\,=\,v\,+\,C[0]\, ,
\eea
with ${\ti v}$ being the renormalized order parameter,  we  arrive
at the following  alternative form  for the order parameter:
\bea \lab{3partsbis} \lan \psi^f(x)\ran &=& {\ti v} \,    + \,  \lf.
{\cal C}[K](x)\ri|_{K=f} \, +\, \lf. D[K](x)\ri|_{K=f}  \, .
\eea
\noindent Note that both $D[K]$ and ${\cal C}[K]$ now vanish when
we set $K=0$,  which is  equivalent to considering the homogeneous
condensation only. In the literature, $C[0]$ is often denoted as $\de
v $ \cite{Coll}.

\subsection{The kink solution in the Born approximation}

\noindent
So far, all result obtained were of a full quantum nature.
We now deal with the
Born, or classical, approximation of Eq.(\ref{3partsbis}) and for this
purpose we reintroduce $\hbar$.
Each propagator
is then proportional to $\hbar$ whilst $B$ has in the exponent the
factor $i\hbar^{-1}$.  The Born approximation means than only terms of
order $\hbar^{0}$ in (\ref{3partsbis}) must be taken into account.
In order to better understand the
classical limit, let us view the field theory as a classical physicist
does: one would then talk about a frequency
$\omega_{0}$ and a wave vector $k$ of the field
rather than about mass $m$ and momentum $p$.
 In QFT, the particle mass and momentum are obtained by multiplying
$\omega_{0}$ and $k$ by $\hbar$ (we do set $c=1$).
The units of $ \lambda = g^{2}/2$  are
$[E^{1}l^{-3}]$ where $E$ means units of energy and $l$ means units of
length, so $[\lambda] = [\hbar^{-1}l^{-2}]$. Note that the latter also
implies a super--renormalizability of the theory.
Similarly, the Fourier transform is expressed in terms of the product
$k x$ and not $p x$. These comments, even if trivial, will be of a
particular importance in the finite temperature case where only a
carefully performed classical limit can provide consistent results. For
instance, in massless theories, a correct classical limit reveals the
breakdown of {\em classical} perturbative expansion (infrared
catastrophe) and then  some resummation is required - the so called
 hard thermal loops (HTL)
resummation\cite{Braaten:1990mz,LB}.

\vspace{3mm}

\noindent In rewriting some of the previous  expressions making
the $\hbar$ dependence explicit, we
recall that the counterterm Lagrangian ${\cal{L}}_{ct}$ is
of order ${\cal{O}}(\hbar)$ and so in the expansion of $C[f]$ and
$D[f]$ terms one always gets loop contributions. It is obvious
that in doing classical limit such contributions must decouple and
so in the following reasonings we may neglect ${\cal{L}}_{ct}$. We
have \footnote{After factorizing out $\hbar$, we shall in the
following consider $\De_{C}(\ldots) \propto \hbar^{0}$.}

\bea C[K](x,\hbar) &=&
\hbar \int_{C} d^{2}y \, \De_{C}(x;y)
\frac{\de}{\de K(y) } \,\exp\lf[\hbar a \ri]
\exp\lf[\frac{1}{\hbar} b\ri]
\\
D[K](x,\hbar) &=& K(x)\,\exp\lf[\hbar a \ri]
\exp\lf[\frac{1}{\hbar} b\ri] \, ,
\eea
\noindent with
\bea a &=& \frac{1}{2}\int_{C}d^{2}zd^{2}y \,\De_{C}(z;y)\,
\frac{\de^{2}}{\de K(z) \de K(y)} \\
b &=&-\frac{i}{2}\int_{C}d^2z\,\lf [
\frac{g^2}{4}\,K^4(z)\,+\, \omega_{0} g\,K^3(z)\ri]\, .
\eea
\noindent Keeping only the finite terms in the $\hbar \rar 0$ limit we
get\footnote{Throughout this paper $\hbar \rar 0$ is a short--hand
notation for $\hbar \rar 0_{+}$.}
\begin{eqnarray} C[K](x,\hbar \rar 0) \non
&=& \int_{C}d^{2}y \,\De_ { C}(x;y)\frac{\de}{\de K(y)}\,
\mbox{Res}_{\hbar =0} \lf ( \exp\lf[\hbar a \ri]
\exp\lf[\frac{1}{\hbar} b\ri]\ri) \non
\\ && \non \\
&=&  \int_{C} d^{2}y \De_{C}(x;y) \,\frac{\de}{\de K(y)}\;
\sum_{n=0}^{\infty} \frac{1}{n!(n+1)!}\, a^{n}b^{n+1} \non
\\ && \non \\
&=& i\int_{-\infty}^\infty d^{2}y\, G_{R}(x,y) \, \frac{\de}{\de
K(y)}\sum_{n=0}^{\infty} \frac{1}{n!(n+1)!}  \, a^{n}b^{n+1} \, ,
\label{CK}
\end{eqnarray}
\noindent where
$iG_{R}(x,y)
=\theta(x_{0}-y_{0})\, \Delta(x;y)$ is the (position space) retarded
Green's function of the free theory and
$\De(x;y)$ is the Pauli--Jordan function:
\bea\lab{pjordan} \De(x_{0},x_{1};0,0) = \lan 0|
[\rho_{in}(x),\rho_{in}(0)] | 0 \ran = \int \frac{d^{2}k}{(2\pi)}\,
\de(k^2 - \omega_{0}^2)\,\varepsilon(k_{0})\, e^{-ikx} = -\frac{i}{2}
\,\theta(x_{0} - |x_{1}|)\, J_{0}(\omega_{0}\, \sqrt{x_{0}^2-
x_{1}^2})\, . \eea

\noindent
The $D$ term gives
\bea \lab{DK} D(x,\hbar \rar 0) &=&\, K(x)\,Res_{\footnotesize
\hbar = 0} \lf( \frac{1}{\hbar} \, \exp\lf[\hbar a \ri]
\exp\lf[\frac{1}{\hbar} b\ri] \ri)
\,=\,K(x)\,\sum_{n=0}^{\infty}\,\frac{1}{(n!)^{2}}\,a^{n}b^{n}\, .
\eea
In Appendix A we show that $D[K, \hbar \rightarrow 0]=K$. The
final result is then
\bea
 \lan \psi_{0}^{f}(x)\ran &= & \lf .  v\,+\, f(x) \,
+ i\,\int_{-\infty}^\infty d^{2}y \, G_{R}(x,y) \, \frac{\de}{\de
K(y)}\; \sum_{n=0}^{\infty} \frac{1}{n!(n+1)!}  \, a^{n}b^{n+1}
\ri|_{K=f}\nonumber  \\
&=& \,v\,+ \, \lf .\sum_{n=1}^\infty P_n[K](x) \ri|_{K=f} \, ,
\label{pj1}
\eea
where
\bea\non
 P_1(x)&=&K (x)
 \\  P_n(x) &=& \,\frac{1}{[(n-2)!]^2}
\int_{-\infty}^\infty d^{2}y\, G_{R}(x,y) \; a^{n-2} \lf\{
\lf[\frac{3}{2}\omega_{0}g \;K^2(y) \,+\, \frac{1}{2} g^2 \; K^3(y)
\ri]\, b^{n-2}\ri\} \, ; \,\,\,\, n\ge 2 . \eea
In Appendix C we show (using mathematical induction) that the
following recurrence relation holds:
\bea P_n(x) \, &=&\, \int_{-\infty}^{\infty}d^{2}y \, G_{R}(x,y)
\;\lf[ \frac{3}{2}\omega_{0} g \;\sum_{i+j=n} \, P_i(y)\, P_j(y)
\,+\, \frac{1}{2} g^2 \sum_{i+j+k=n+1} \, P_i(y)\, P_j(y)\, P_k(y)
\quad \ri] \, ; \,\,\,\, n\ge 2 , \lab{44} \eea
where $i,j,k =1,2 \ldots$. The recurrence condition (\ref{44}) may
be ``diagonalized'' in the following way
\begin{eqnarray}
Q_1(x)&=&K (x) \nonumber \\
Q_2(x)&=& \int_{-\infty}^{\infty}d^{2}y \frac{3}{2v}\,
G_{R}(x,y)\,
\;\sum_{i+j=2} \, Q_i(y)\, Q_j(y)  \nonumber \\
Q_n(x) &=& \int_{-\infty}^{\infty}d^{2}y\, G_{R}(x,y)\;\lf[
\frac{3}{2}\omega_{0} g \;\sum_{i+j=n} \, Q_i(y)\, Q_j(y) \ri.
+\,\lf.  \frac{1}{2} g^2 \sum_{i+j+k=n} \, Q_i(y)\, Q_j(y)\,
Q_k(y) \ri] \, ; \,\,\,\, n\ge 3\, . \label{pj45}
\end{eqnarray}
\noindent with
\bea
Q_{n} = P_{1}P_{n+1} + P_{n}\, .
\eea
\noi Note that both $\sum_{i=1}^{\infty}P_{i}$ and
$\sum_{i=1}^{\infty} Q_{i}$ lead to the same result\footnote{This might
be immediately seen from the fact that both $\sum_{i=1}P_{i}$ and
$\sum_{i=1}Q_{i}$ solve the same integral equation, namely
\begin{displaymath}
X(x)-f(x) = \int_{-\infty}^{\infty}d^{2}y\, G_{R}(x,y)\, \left[
\frac{3}{2}\omega_{0}g X^{2}(y) +
\frac{1}{2}g^{2}X^{3}(y)\right]\, .
\end{displaymath}
}and so it does not really matter which recurrence equation will be
solved.  The formulation of the problem in terms of recurrence
conditions (\ref{pj45}) has however one crucial advantage, namely
Eq.(\ref{pj45}) belongs to the class of the so called functional
equations of Cauchy--Marley's type\cite{Bell,Mar} where it is known
that the fundamental solution cannot be expressed (apart from a very
narrow class of kernels) in terms of elementary functions. This means
that we cannot resolve (\ref{pj45}) in terms of general $K$ (or
$f$), i.e. we cannot find the B{\"{a}}cklund
transformation between the solutions of Eq.(\ref{cpt3b})  and
those of Eq.(\ref{euler}) - the classical Euler-Lagrange equation of motion
(see below). Nevertheless we can  obtain all the
analytical solutions of Eq.(\ref{euler}),namely the analytical kinks:
 This is done once we realize that the convolution of the $2D$
retarded Green's function $G_{R}(x)$ with an exponential is
proportional to the very same exponential (actually the exponential is
the only function having this property - see Appendix C). So we choose
$Q_{n}(x) \propto (Q_{1}(x))^{n}$ with $Q_{1}(x) = f_{0}(x)$ being an
exponential, Fourier non--transformable solution of the  Klein--Gordon
equation (\ref{cpt3b}). We obtain:
\bea
Q_{n}(x) = A_{n}f^{n}_{0}(x) =  A_{n}\, e^{\pm \omega_{0} n\gamma(x_{1}
-x_{0}u)}\, ,
\eea
\noindent where $\gamma = (1-u^2)^{-\frac{1}{2}}$ ($u$ will be
later interpreted as a velocity of a kink). Plugging this form
into the recurrence relation (\ref{pj45}) and using the result
from Appendix C we arrive at the following equation for the factor
$A_{n}$
\begin{equation}
A_{n}= \frac{1}{(n^{2}-1)}\left\{\frac{3}{2v}\sum_{i+j=n} A_{i}A_{j} +
\frac{1}{2 v^{2}}\sum_{i+j+k=n} A_{i}A_{j}A_{k}\ri\}\, .
\label{pj2}
\end{equation}

\noi This is a trivial version of Cauchy--Marley's equation which has
the only (non--zero) fundamental solution, namely $A_{n} \propto
(A_{1})^{n}$.  Using the standard identities:

\bea
\sum_{i+j=n}1 &=& n-1 \qquad ; \qquad
\sum_{i+j+k=n}1 \,=\, \frac{1}{2}(n-1)(n-2)\, ,
\eea
\noindent
it might be easily checked that a solution of the recurrence relation
(\ref{pj2}) reads
\bea \lab{ansatz} A_n\, =\, 2 v \, \lf(\frac{s}{2v}\ri)^n\, , \eea

\noindent with $c$ being a real constant. Thus, finally, we have
\bea \lan \psi_{0}^{f}(x)\ran \, &=& \, v +
2v\sum_{n=1}^{\infty}\lf(\frac{s f_{0}(x)}{2v}\ri)^{n}  \;=\;  v\,
\mbox{cth}\lf[\frac{1}{2} \mbox{Ln} \lf(\frac{s
f_{0}(x)}{2v}\ri)\ri]\, .
\lab{sol1} \end{eqnarray}
\noindent Here  $\mbox{Ln}(z) = \mbox{ln}|z| \,  + \, i\,\mbox{arg}z$
is the usual principal value of the logarithm of $z$. Note that $K =
f = sf_0$.  Thus, provided
$f(x)$ is an exponential solution of the linear equation
(\ref{cpt3b}), the solution (\ref{sol1}) fulfils the (classical)
Euler--Lagrange equation of motion:
\bea\lab{euler} (\pa^{2} + \mu^{2})\lan \psi_{0}^{f}(x)\ran \, =\, -
\la \,\lan \psi_{0}^{f}(x)\ran^{3} \qquad,\qquad \mu^{2} =
- \omega_{0}^{2}/2 \, . \eea
\noi The later is nothing but the expectation value of equation
(\ref{pj34}) in the Born approximation.

\vspace{3mm}

\noindent For instance, if we choose  $f_{0}(x) =
e^{-\omega_{0}\gamma(x_{1} -x_{0}u)}$ with $s=-2v e^{\omega_{0}\gamma
a}$, we easily obtain the standard kink solution\cite{Da,Go,Po}
\bea \lan \psi_{0}^{f}(x) \ran \,= \,v \;
\mbox{th}\lf[\frac{\omega_{0}}{2}\gamma ((x_{1}-a) -x_{0}u)\ri]\, ,
\eea
describing a constantly moving kink of a permanent
profile with a center localized at $a+ux_{0}$. Note that the
former function $f(x)$ is the solution of the homogeneous
Klein--Gordon equation (\ref{cpt3b}), is not Fourier transformable
and fulfils the initial value condition: $f(x_{0}\rar
-\mbox{sgn}(u)\infty, x_{1})\rar 0$.

\vspace{3mm}

\noindent As the Lagrangian (\ref{lagr1}) is $Z_{2}$ invariant we
could equally choose $\langle 0|\psi(x) | 0 \rangle = -v$. In this
case we would get
 \bea \lan \psi_{0}^{f}(x) \ran \,= \,-v \;
\mbox{th}\lf[\frac{\omega_{0}}{2}\gamma ((x_{1}-a) -x_{0}u)\ri]\,, \eea
\noindent which is the antikink solution. Note
that the antikink can be also alternatively obtained choosing $f_{0}(x)
= e^{+\omega_{0}\gamma(x_{1}-x_{0}u)}$  and $s= -2ve^{-\omega_{0}\gamma
a}$, provided we keep  $\langle 0|\psi(x) | 0 \rangle = v$.

\vspace{3mm}

\noindent It is quite instructive to realize what happens if we choose
$s$ positive, i.e. $s = 2v e^{\omega_{0} \gamma a}$. In this case we
obtain
\bea\lan \psi_{0}^{f}(x) \ran \,= \,\pm v \;
\mbox{cth}\lf[\frac{\omega_{0}}{2}\gamma ((x_{1}-a) -x_{0}u)\ri]\, ,
\eea
\noindent ($\pm$ sign depends on the choice of $f_{0}(x)$) which is also
solution of Eq.(\ref{euler}). However, this solution is singular in
$x_{1}-a = x_{0}u$ and so it does not classify as a soliton even if,
for example, the corresponding energy density

\bea \non
{\cal{H}}_{cl}(x) &=& \,\frac{1}{2}(\langle \psi^{f}_{0}(x){\dot
{\rangle}})^{2} + \frac{1}{2}(\langle \psi^{f}_{0}(x)\rangle ')^{2} -
\frac{1}{4}\omega^{2}_{0}\langle \psi^{f}_{0}(x) \rangle^{2} +
\frac{\lambda}{4}\langle \psi^{f}_{0}(x) \rangle^{4} +
\frac{1}{8}\omega_{0}^{2}v^{2} \\[2mm]   &=&  \,
\frac{1}{4(1-u^{2})}v^{2}\omega_{0}^{2}\, \mbox{csch}^{4} \left[
\frac{\omega_{0}}{2}\gamma ((x_{1}- a) - x_{0}u)\right] \, ,
\eea

\noindent is localized near $x_{1}= x_{0}u +a$, and the corresponding
total energy

\bea
E_{cl}= \int_{-\infty}^{\infty} dx_{1} \; {\cal{H}}_{cl} (0,x_{1})\,
=\, \frac{\omega^{3}_{0}}{3 \lambda} \gamma \, =\, M_{cl}\gamma\, ,
\eea

\noindent is finite ($M_{cl}$ denotes the classical (static) kinkmass).

\vspace{3mm}
\noindent
We thus see  that, in the classical limit, the present method gives us a
constructive way for
finding classes of analytical solutions of non--linear
equations. The fact that we could not find the fundamental solution of
Eq.(\ref{cpt3b}) is intimately connected with the theory of
Cauchy--Marley's functional equations. On the other hand the
non--existence of the fundamental solution is fully consistent with
the fact that  $(1+1)$ dimensional $\lambda \psi^{4}$ theory is not
exactly solvable.
\vspace{3mm}

\noi The above solutions for the shift function $f$ corresponding
to kink solutions
 can be now used in Eq.(\ref{cpt72}) to get the Heisenberg
operator in a given kink sector. One can thus calculate Green's functions  and
higher quantum corrections to $\phi^f$. We show this elsewhere
\cite{Blasone:1999aq,toppress}.

\subsection{The kink solution in the small coupling limit}

\noindent We can obtain the classical limit of the
order parameter (\ref{3partsbis}) using a  different standpoint,
namely the semiclassical (or WKB) approximation \cite{landau}.
This asserts that a system  behaves classically if its de Broglie
wavelength $\lambda_{dB}$ is much smaller  than the characteristic
length scale $r_{0}$ of the theory: at this stage  the wave
properties of a system become unimportant. In our case the kink
(antikink) de Broglie wavelength is $\lambda_{dB} =  2\pi \hbar /
|p| = 2\pi\hbar / \gamma M_{cl} |u|$. On the other hand, $r_{0}$
in the theory can be identified with the kink (antikink)
width ($= 1/\gamma\omega_{0}$). So
\begin{equation}
\lambda_{dB} \, \ll \, r_{0} \; \; \; \Rightarrow \; \; \; \hbar
\lambda \, \ll \,  \omega_{0}^{2} |u| \; \; \; \Leftrightarrow \; \;
\; \hbar g^{2} \ll \omega^{2}_{0} |u|\, .
\label{dB1}
\end{equation}
\noindent Note that the above condition  can be fulfilled in all
Lorentz frames of reference apart from the ones where  the kink
(antikink) is close to or exactly at rest. This has a simple
physical meaning: the semiclassical approximation is based on the
assumption that typical momentum and distance scales are much
larger than the error scales saturating Heisenberg's uncertainty
relations. This assumption naturally fails when the typical
momentum scale is close to zero and correspondingly for small
$|u|$ the semiclassical approximation cannot be used.

\vspace{3mm}

\noindent Eq.(\ref{dB1}) is fulfilled either if $\hbar \rightarrow
0$ (see previous subsection) or if $\lambda \rightarrow 0$, i.e.
in the heavy mass approximation. We study the latter case in
Appendix D: there we show  that  the small coupling limit,
leads to the analytical kink (antikink) solutions of the classical
Euler--Lagrange equation (\ref{euler}). We stress once again that
the classical soliton solutions obtained both in the previous and
present subsections were determined as a classical limit of the
{\em full} quantum mechanical expression for the order parameter
(\ref{3partsbis}).

\section{Kinks in two--dimensional $\la \; \psi^{4}$ theory at $T\not=0$}

\noindent So far we have dealt with QFT  at zero
temperature.  However, in many physical situations,
the zero temperature approximation is not appropriate. This
happens for example in cosmology\cite{CWbe,LDo,ALi,RGLe},
astrophysics\cite{TAl,EBr,EBr2} and in the study of
the quark--gluon plasma
formation\cite{RBa,TAl2,EBr3} as well as in condensed matter.
More in general, in all cases when one studies systems near
the critical temperature   (and so neither low  or high temperature
expansions properly describe the basic features) the finite--temperature
treatment must be carefully taken into account.

\vspace{3mm}

\noindent  It is important therefore to understand what happens with the
topological defects when the QFT system is immersed in a heat bath
(thermal reservoir) which is saturated at some temperature $T$. In this
Section we aim to use our toy--model system to address this question and
demonstrate the basic features of our approach in the
finite--temperature case. Most of the results here obtained retain their
validity when more realistic (higher dimensional) systems are considered.

\vspace{3mm}

\noindent
As already discussed in Section II.B, in thermal equilibrium, the most
convenient choice for the time path is the one in Fig.2, to which we
refer as the {\em thermal path}.
The crucial observation at finite temperature is that the
operatorial Wick's theorem still holds (see e.g. \cite{Evans:1996bh})
and
consequently Eq.(\ref{cpt72}) retains its validity provided the
following substitutions are performed:
\begin{eqnarray*}
\Delta_{C}(x;y) = \langle 0| T_{C}(\rho_{in}(x) \rho_{in}(y)) | 0
\rangle \; &\longrightarrow & \; \Delta_{C}(x;y,T) = \langle T_{C}(\rho_{in}(x)
\rho_{in}(y)) \rangle_{\beta}\\ [2mm] :\ldots :\; &\longrightarrow& \;
N(\ldots )\, ,
\end{eqnarray*}
\noindent where $\langle \ldots \rangle_{\beta} \equiv
Tr\left(e^{-\beta H}  \ldots \right) /\mbox{Tr}\left( e^{-\beta H}
\right) $  and $\beta = 1/T$.  The thermal normal ordering $N(\ldots)$
is defined is such a way\cite{Evans:1996bh}  that $\langle N(\ldots )
\rangle_{\beta} = 0 $, the dots stands for a product of $T=0$ free
fields. This is of a great importance as all the formal considerations
developed in Section III.A go through also for finite $T$.

\subsection{The kink solution at finite $T$ in the classical approximation}

\noindent The classical approximation has proved to be a useful
tool for study of low--energy and/or high--temperature properties
of quantum fields at finite temperature. Examples include
 non--perturbative computations of the Chern--Simons
diffusion rate\cite{Moo1}, sphaleron decay\cite{Bo1} or plasmon
properties of hot non--abelian gauge theories\cite{TS1}; there is
also a perturbative treatment
via the {\em classical} Feynman diagrams\cite{GA1} which  however will
not be followed here.

\vspace{3mm}

\noindent At nonzero temperature the question of $\hbar$ appearance is
more delicate than in the zero--temperature case. The whole
complication is hidden in the thermal propagator
$\Delta_{C}(x;y,T)$. Whilst at $T=0$ the latter is directly
proportional to $\hbar$, at finite $T$ the situation is very
different. Although this generic observation appears simple, there has
been in the past confusion in the literature about this point which
was understood properly only recently \cite{Braaten:1990mz,Blaizot:1993gn}.
To understand the complications involved let us make
$\hbar$ explicit. The free thermal propagator in spectral or Mills's
representation\cite{Mills,LB} then reads
\begin{eqnarray}
\Delta_{C}(x;y,T) &=& \hbar\int\frac{d^{2}k}{(2\pi)^{2}} \,
e^{-ik(x-y)}\,\rho(k) [\theta_{C}(x_{0}-y_{0}) + f_{b}(\hbar k_{0}/T)
]\nonumber \\ &=& \Delta_{C}(x;y) + \Delta^{T}_{C}(x;y) \, ,
\lab{prop2}
\end{eqnarray}
with
\bea
\Delta_{C}(x;y) = \hbar \int
\frac{d^{2}k}{(2\pi)^{2}}e^{-ik(x-y)}\rho(k) \left[ \theta_{C}(x_{0}-
y_{0}) - \theta (-k_{0}) \right]\, ,
\eea
where the spectral density $\rho(k) =
(2\pi)\varepsilon(k_{0})\, \delta(k^{2}-\omega_{0}^{2})$ with
$\varepsilon(k_{0}) = \theta(k_{0}) - \theta(-k_{0})$. The contour
step function $\theta_{C}(x_{0}-y_{0})$ is $1$ if $y_{0}$ precedes
$x_{0}$ along the contour $C$. The Bose--Einstein distribution
$f_{b}(x)= \left(e^{x}-1\right)^{-1}$.  It might be directly checked
that we obtain the usual elements of the (free) thermal propagator in
the so called Keldysh--Schwinger formalism \cite{Landsman:1987uw,LB}, i.e.:
\begin{eqnarray} \non
\Delta_{11}(x;y,T) &=& \hbar \int \frac{d^{2}k}{(2\pi)^{2}} \,
e^{-ik(x-y)}\left\{ \frac{i}{k^{2}-\omega^{2}_{0} + i\varepsilon}    +
2\pi \, \delta(k^{2}-\omega_{0}^{2})f_{b}(\hbar |k_{0}|/T)
\right\}
\\ \lab{matrprop}
 \Delta_{21}(x;y,T) &=& \hbar \int
\frac{d^{2}k}{(2\pi)}\, e^{-ik(x-y)}\left\{ \theta(k_{0})  +
f_{b}(\hbar |k_{0}|/T)\right\} \delta(k^{2} - \omega_{0}^{2})
\\ \nonumber
 \Delta_{22}(x;y,T) &=& (\Delta_{11}(x;y,T))^ {*}
\\ \non
\Delta_{12}(x;y,T) &=& (\Delta_{21}(x;y,T))^{*}\, ,
\end{eqnarray}
\noi where $``^{*}"$ denotes complex conjugation and the particle mass $m
= \hbar \omega_{0}$. We remark that the thermal part of
$\Delta_{C}(x;T)$ is identical for all matrix elements and that
in (\ref{matrprop}) it appears $f_{b}(\hbar |k_{0}|/T)$ and not
$f_{b}(\hbar k_{0}/T)$. We should also emphasize that $k$ in the
integration is a wave vector - a reciprocal length - and not a
momentum.

\vspace{3mm}

\noindent Now, due to the mentioned analogy with the $T=0$ situation,
we may immediately write for the order parameter
\begin{eqnarray}
\lan \psi^f(x)\ran_\bt &=& v \, +\, C[K](x;T)|_{K=f} \,+\,
D[K](x;T)|_{K=f} \, , \label{B233}
\end{eqnarray}
\noi where  we took the thermal average of the expression
analogous to the one in Eq.(\ref{cpt72}), but with the normal
ordering and the propagator replaced with their thermal
counterparts. Both  $C[K](x;T)$ and
$D[K](x;T)$ entering Eq.(\ref{B233}) coincide with their zero
temperature counterparts provided one uses the thermal propagator
instead of $\Delta_{C}(x,y)$.

\vspace{3mm}

\noindent Let us consider the classical limit of Eq.(\ref{B233}).
The $D[f](x;T)$ term  then gives:
\begin{eqnarray}
D[f](x;T, \hbar \rightarrow 0) &=& f(x)\,Res_{\hbar = 0} \left[
\frac{1}{\hbar} \exp\left(\frac{\hbar}{2}\int_{C} d^{2}z d^{2}y
\, \Delta_{C}(z;y,T) \frac{\delta^{2}}{\delta K(z)\delta K(y)}
\right)\right.\nonumber \\
&& \times \, \left. \left .\exp \left(-\frac{i}{2\hbar} \int_{C}
d^{2}z \, \left[ \frac{g^{2}}{4} K^{4}(z) + \omega_{0}g
K^{3}(z)\right]\right)\right]\right|_{K=f}\nonumber \\ &=& \left.
f(x)\, Res_{\hbar = 0} \sum_{n,m=0}  \left[
\frac{\hbar^{n-m-1}}{n!\, m!} \left( \frac{1}{2} \int_{C} d^{2}z
d^{2}y \, \Delta_{C}(z;y,T) \frac{\delta^{2}}{\delta K(z) \delta
K(y) } \right)^{n}b^{m} \right]\right|_{K=f}\, . \label{D11}
\end{eqnarray}
\noindent Note that if $\Delta^{T}_{C}(x;y) = 0$ we recover the
result (\ref{DK}) of the previous section. Using the result of
Appendix A we may directly write
\begin{eqnarray}
D[f](x;T, \hbar \rightarrow 0) &=& f(x) \, Res_{\hbar = 0}
\sum_{n,m}\left[ \frac{\hbar^{n-m -1}}{n!\,m!}\, \delta_{n0}\,
\delta_{m0} \right] \, = \, f(x)\, . \label{B336}
\end{eqnarray}

\noindent For the  $C[f](x;T)$ term we get, in the classical limit:
\begin{eqnarray}
&&C[f](x;T, \hbar \rightarrow
0)\nonumber \\
&&\mbox{\hspace{1cm}} = \left.\int_{C}d^{2}y \, \Delta_{C}(x,y)
\frac{\delta}{\delta K(y)}\, Res_{\hbar \rightarrow
0}\sum_{n,m}\left[ \frac{\hbar^{n-m}}{n!\, m!} \left(
\frac{1}{2} \int_{C}d^{2}z d^{2}w \,
\Delta_{C}(z;w,T)\frac{\delta^{2}}{\delta K(z) \delta K(w)}
\right)^{n} b^{m}\right]\right|_{K=f} \label{B337}
\end{eqnarray}
\noindent There is no simple way how to evaluate
(\ref{B337}) without performing an explicit Laurent's expansion of
$\Delta_{C}(x,T)$ around $\hbar = 0$. Using the Bernoulli expansion:
\begin{displaymath}
\frac{x}{e^{x} - 1} = \sum_{\alpha = 0}^{\infty}
B_{\alpha} \, \frac{x^{\alpha}}{\alpha !}\, , \;\;\;\;\;\;\;\;  |x|< 2\pi \, ,
\end{displaymath}
\noindent ($B_{\alpha}$ are Bernoulli's numbers) we may
Laurent expand $f_{b}$ as
\begin{equation}
f_{b}(\hbar k_{0}/T) +
\frac{1}{2} = \frac{T}{\hbar k_{0}} + \frac{1}{12} \frac{\hbar k_{0}}{T} -
\frac{1}{90} \left(\frac{\hbar k_{0}}{T}\right)^{2} + \, \ldots \, .
\label{III45}
\end{equation}

\noindent The key observation is that the foregoing series converges only for
$\hbar |k_{0}| < 2\pi T$. The leading term in (\ref{III45}) gives the {\em
classical} thermal part of the propagator\footnote{We can call it
Rayleigh--Jeans's sector of the thermal propagator as the corresponding
distribution function $f(\omega) = T/\omega$ is nothing but
Rayleigh--Jeans's distribution law.}, and the classical approximation is then
equivalent to taking the leading term in the Laurent expansion
(\ref{III45}). The higher quantum corrections are due to the higher terms
in the expansion, but for large $|k_{0}|$ the expansion does not work,
i.e. an expansion in $\hbar$ is unwarranted. Of course, for $\hbar |k_{0}|
\gg T$ the distribution $f_{b}$ is exponentially small (Wien's
distribution law) and it is dropped in comparison to the (zero point)
first term in the integral (\ref{prop2}), which returns the usual $T=0$
approach from the previous Section.

\vspace{3mm}

\noindent In order to get some quantitative results let us
concentrate on the high temperature case (low temperature case is
more involved and will be treated elsewhere). By {\em high}
temperature we mean the temperature at which the Rayleigh--Jeans's
sector of the thermal propagator (\ref{prop2}) approximates by
itself sufficiently well the thermal part of the propagator. This
should be taken with a grain of salt as the  high temperature we
intend to use should be still bellow the temperature threshold
above which the heat--bath quasiparticles become instable due to
the thermal fluctuations.

\vspace{3mm}

\noindent Because we are in the high temperature regime we
take cut off in the $k_{0}$ integration: $ |k_{0}| \ll \Lambda_{1}
\approx T/\hbar$. Due to $\delta$--function in the thermal
propagator we must also impose cut off on $k_{1}$ integration:
$|k_{1}| \ll \Lambda_{2}\approx T/\hbar$. We tacitly assume that
the domain of integration from $T/\hbar$ to $\infty$ will
contribute modestly due to the exponentially small contribution to
the thermal propagator.

\vspace{3mm}

\noi Using (\ref{CK}), (\ref{sol1}) and (\ref{B337}) we can now
evaluate $C[f](x;T, \hbar \rightarrow 0)$. Indeed
\begin{eqnarray}
C[f](x;T, \hbar \rightarrow 0) &=& \lf.\exp\left[ c \right]\,
\left\{ i \int_{-\infty}^{\infty} d^2y \, G_{R}(x,y)
\frac{\delta}{\delta K(y)} \, \sum_{n=0}^{\infty}
\frac{1}{n!(n+1)!} a^n b^{n+1}\right\}\ri|_{K = f} \nonumber \\ [2mm]
&=& \lf.\exp\left[ c \right]\,\left\{-K(x) -2v -\frac{4v^2}{K(x) -
2v} \right\}\ri|_{K=f}\, , \label{cth1}
\end{eqnarray}
\noi where
\begin{eqnarray}
c &=& \frac{\hbar}{2}\, \int_{C}d^2zd^2y \, \Delta_C^T(z;y)
\frac{\delta^2}{\delta K(z)\delta K(y)} \, .
\label{c22}
\end{eqnarray}
\noi Note that there is no $\hbar$ in $c$ in the large $T$
approximation.  In other words, the high temperature properties are injected
into the order parameter calculation in a purely classical way.
This might be compared with
the usual observation that at large $T$ thermal fluctuations
dominate over quantum ones. Thus in our case we get
(we omit $\hbar$ in $\Delta_C^T$
as it cancels anyway in the large $T$
limit)\footnote{ To derive Eq.(\ref{C11})
we use the relation:
$\,\exp\left(\sum_{ij} A_{ij} \partial_{x_i}\partial_{x_j} \right) \,
F(x_k) = F(x_k + 2 \sum_j A_{kj}\partial_{x_j}) \ident\, .$
}
\begin{eqnarray}
C[f](x;T, \hbar \rightarrow 0) = \left.\left\{-K(x)-2v
-\frac{4v^2}{ K(x) + \int_C d^2y \Delta_C^T(x;y)
\frac{\delta}{\delta K(y)}   - 2v} \ident \right\}\right|_{K=f}\,
. \label{C11}
\end{eqnarray}
\noi Applying the expansion
\begin{displaymath}
\frac{1}{A+B} = A^{-1} - A^{-1} B A^{-1} + A^{-1}BA^{-1}BA^{-1} - \ldots\, ,
\end{displaymath}
\noi and identifying $A = K(x) - 2v $ and $B =  \int_C d^2y \ \Delta_C^T(x;y)
\frac{\delta}{\delta K(y)} $ we get
\begin{eqnarray}
C[f](x;T, \hbar \rightarrow 0) &=& - f(x) -2v - 4v^2
\left(\frac{1}{K(x) - 2v} + \lf.\frac{ \Delta_{11}^{T}(0)}{ (K(x) -
2v)^3} + 3 \ \frac{[\Delta_{11}^{T}(0)]^2}{( K(x) - 2v)^5} + \ldots
\right)\ri|_{K=f}\nonumber \\
&&\nonumber \\
&& = - f(x) -2v - \frac{4v^2}{f(x) - 2v} - \left. 4v^2\sum_{n=1}^{\infty}
\frac{ [\Delta_{11}^{T}(0)]^n \, }{(K(x) - 2v)^{2n+1}} \;
(2n-1)!! \, \right|_{K=f}\, .
\label{er1}
\end{eqnarray}
\noi The series $\Sigma(z) = 1 + \sum_{n=1}^{\infty} z^n (2n-1)!! \equiv
\sum_{n=0}^{\infty} \sigma_{n} z^n$ is clearly divergent for all
non--trivial values of $z = \Delta_{11}^{T}(0)/(K(x) - 2v)^2$.  Thus
$\Sigma(z)$ can be at best understood as an asymptotic series. Question
arises then which function has $\Sigma(z)$ as its asymptotic expansion.
To determine this, let us assume initially that  $z \in \mathbb{C}$ and
perform the Borel transform  $B_{\Sigma}(z)$ of $\Sigma(z)$ by \cite{Er1}
\begin{equation}
B_{\Sigma}(z) = \sum_{n=0}^{\infty}B_n z^n \equiv
\sum_{n=0}^{\infty} \frac{\sigma_n}{n!} z^n = \frac{1}{\sqrt{1-2z}} \, .
\end{equation}
\noi Consequently the Borel sum of $\Sigma(z)$ is
\begin{equation}
\int_{0}^{\infty}dt \, e^{-t} B_{\Sigma}(tz) = - \sqrt{\frac{\pi}{-2z}} \
e^{-1/2z} \
\mbox{Erfc}\left( \frac{1}{\sqrt{-2z}}\right)\, ,
\label{Bt2}
\end{equation}
%
%
\noi where $\mbox{Erfc}(\ldots)$ is the complementary error
function (see e.g.,\cite{AS1}). Relation (\ref{Bt2}) is true for any
$z \in \mathbb{C}$ with $\arg(z) \not= 0$, and so in such a region
$\Sigma(z)$ is Borel summable with the unique asymptotic function
\begin{equation}
- \sqrt{\frac{\pi}{-2z}} \
e^{-1/2z} \
\mbox{Erfc}\left( \frac{1}{\sqrt{-2z}}\right)
\sim \Sigma(z)\, , \;\;\;\;\; \arg(z) \not= 0\, .
\label{Bt3}
\end{equation}
\noi In the marginal case when $\arg(z) = 0_{-}$ or $\arg(z) = 0_{+}$
we have
\begin{equation}
-i \sqrt{\frac{\pi}{2z}} \, e^{-1/2z} \  \mbox{Erfc}\left( i
\frac{1}{\sqrt{2z}} \right) \sim \Sigma(z)\; \;\;\;
\mbox{or} \;\;\;\;
i\sqrt{\frac{2\pi}{z}} \, e^{-1/z} - i \sqrt{\frac{\pi}{2z}} \,
e^{-1/2z} \ \mbox{Erfc}\left( i
\frac{1}{\sqrt{2z}} \right) \sim \Sigma(z)\, ,
\label{Bt4}
\end{equation}
\noi respectively, depending which value of $\sqrt{-1}$ one  accepts
along the branch cut.  From the uniqueness of the asymptotic expansion
(\ref{Bt3}) follows    that when $z$ is real, then the functions in
(\ref{Bt4}) are the only ones which have $\Sigma(z)$ as their asymptotic
expansion. This dichotomy is inherent in theory of asymptotic
expansions\cite{Er1} and fortunately is not counterproductive
here. Both functions differ by an exponent which rapidly vanishes when
$z \rightarrow 0$ and accordingly they belong to the same  equivalence
class of asymptotic functions (see e.g., Poincar\'{e}
criterion\cite{Er1}). A brief inspection of (\ref{Bt4}) reveals an
important feature of the class functions, namely that they are complex even
when $z$ is real.
\vspace{3mm}

\noi Applying now the fact that in ($1+1$) dimensions the thermal tadpole
$\Delta_{11}^{T}(0)= T/2\omega_0 = T/2gv$, we see that there is no high
$T$ solution for the thermal kink as the order parameter turns out to be
a complex number. This result is fully compatible with the absence of
phase transitions in ($1+1$) dimensions\cite{MW1,Co1}. In fact, the
emergence of the complex valued order parameter can be attributed
(similarly as in the case when the effective action machinery is in use)
to spontaneous fluctuations in the order parameter. The latter behavior
is well known, for instance, from Ising--type models\cite{par1} or from
numerical simulation in $(1+1)$ dimensional  $\phi^4$
theories\cite{laguna}.

\vspace{3mm}

\noi It is interesting to compare the above result with the one
contained in Ref.\cite{vitman}: there a tree approximation at finite
temperature is used to get a ``classical" kink solution which is of the
same form as the zero--temperature one but with thermal parameters
$m(T)$ and $\lambda(T)$ in the place of zero--temperature ones. From our
discussion it is clear that the tree approximation is not useful in
extracting classical results at finite temperature as the loopwise
expansion is not anymore an expansion in $\hbar$.  In fact, a
resummation of  infinitely many (thermal) loops is indeed
necessary\footnote{It should be noted that the paradigmatic example of
this fact is HTL resummation  which
is used to cure the breakdown of the conventional perturbative theory
for infrared momenta in thermal field theories with light bosons
(especially in gauge theories where symmetries prevent perturbative
radiative generation of masses)\cite{Braaten:1990mz,Blaizot:1993gn}.}:
the resummation in (\ref{er1})
takes neatly care of this.  However, the qualitative result of
Ref.\cite{vitman} remains valid: at some ``critical" temperature the
kink solution disappears. This is not evident in the present case
because of the dimensionality (reflecting in the temperature dependence of the
propagator),  nevertheless the above analysis  till first line of
Eq.(\ref{cth1}) is general and can be applied, for example,
 to  $\la \psi^4 $ in
($3+1$) dimensions: there, in the thin wall approximation,
analytic kink solutions with spherical symmetry have essentially the
same form as in ($1+1$) dimensions \cite{linde}.  It is definitely
an intriguing
question if in this case the temperature at which the kink disappears
coincides or not with the critical temperature at which the symmetry is
restored.  Note that both of these temperatures can be estimated in the
present approach: for instance,  the one for symmetry restoration can be
read from the thermal order parameter  in the case of  homogeneous
condensation ($K=0$).  Work in this direction is currently in
progress\cite{toppress}.

\section{Conclusions and outlook}

\noi In this paper we have shown how to describe topological defects as
inhomogeneous condensates in Quantum Field Theory, both at zero
and at finite temperature. To this end, we have used the
Closed--Time--Path formalism, allowing us to obtain a closed functional
expansion for the Heisenberg field operator in presence of
defects: the inhomogeneity of the vacuum condensate is introduced by means of
a ``shift function'' $f(x)$ which is solution of the equations
for the physical fields (quasiparticles).

\vspace{3mm}

\noi We applied this general scheme to a specific simple model - (1+1)
dimensional $\la \psi^4$ theory. We constructed the Heisenberg
operator $\psi^f$ in the kink sector, showing explicitly how the
known classical kink solutions arise from the vacuum expectation
value of such an operator in the classical limit. The knowledge of
$\psi^f$ gives us the possibility of  calculating
 quantum corrections to the soliton
solutions as well as Green's functions in presence of kinks.

\vspace{3mm}

\noi We have also shown how to extend this treatment to finite
temperature and  considered explicitly the case of high
temperature. In this limit, we are able to calculate the order
parameter in the classical limit, showing how kink solutions
depend on temperature. This is of particular interest in the case
of higher dimensionality: then  the analysis presented here
goes through in a similar way and it is possible to study the
behavior of topological defects near the critical temperature, as
well as to calculate $T_c$. In fact, when the case of homogeneous
condensation ($f=0$) is considered,  our formulation offers a way
to the study of restoration of symmetry, alternative to
the traditional analysis based on  effective potentials.

\vspace{3mm}

\noi Natural extensions of the present work includes the sine--Gordon
model and the (3+1) dimensional $\la \psi^4$ theory.  For the
sine--Gordon model, which is integrable,
it is of interest to investigate the
correspondence between the present  method and  the Backl\"und
transformations.  We have indeed shown that in the classical limit, the
introduction of the shift function $f$ allows for  a linearization of
the equation for the order parameter. Then for integrable systems,
 linear superposition of
solutions of the (linear) equation for $f$ give rise to multisoliton
solutions for the non-linear Euler equation: such a correspondence could
indeed offer an alternative possibility for classifying soliton solutions
(i.e. by means of the shift function $f$).
On the other hand,
we have explicitly shown that, in the case of (1+1) dimensional
$\la \psi^4$ (which is non--integrable), it is not possible
to solve in general the recurrence relation establishing the non--linear
to linear mapping, except for a restricted class of solutions for
$f$.
Other interesting aspects
of the sine--Gordon model for which the present approach could be
useful are the finite temperature behavior \cite{Lu:2000iy} and the
duality with the Thirring model both at zero \cite{Coleman} and finite
temperature \cite{Steer:1998ah}.

\vspace{3mm}

\noi For the  (3+1) dimensional $\la \psi^4$
theory with complex field - relativistic Landau--Ginzburg theory,
we already found an interesting relation between the Green's functions for
the system with vortices and the one without vortices
\cite{Blasone:1999aq}, allowing us to calculate a
``topological" contribution to the hydrostatic pressure
\cite{toppress}. Temperature--induced
restoration of symmetry is an another
important aspect which can be studied in the present framework.

\vspace{3mm}

\noi Finally, it is definitely a challenging task to extend the above
formulation to the description of phase transitions in which
defects could be created. The main problem in doing this is in the
change of the set of the asymptotic fields when crossing different
phases, which is not taken into account in the present
formulation. Study in this direction is in progress.




\vspace{3mm}

\noindent

\section*{Acknowledgments}
\noi We acknowledge partial support from EPSRC, INFN,  Royal
Society and the  ESF network ``COSLAB".

\section*{Appendix A}

\noindent We show here that $D[f, \hbar \rightarrow 0] = f$. To
prove this we start with the definition of $D[K,\hbar \rightarrow
0]$. Let us remind that
\begin{equation}
D[K, \hbar \rightarrow 0 ] =
K(x)\sum_{n=0}^{\infty}\frac{1}{(n!)^{2}}a^{n}b^{n}\, , \lab{B1}
\end{equation}
\noindent with $a,b$ being defined in Section III.A. We shall now show
that $a^{k}b^{l}[K=f]=1$ if $k=l=0$ and $0$ otherwise (both $k$ and $l$
are positive integers). In the case when $k=l=0$ or $l=0, k\not=0$  our
statement obviously holds. If $k=0$ but $l\not= 0$ our statement is true due
to the fact that the contour integration of the function which is
continuous across the real--time axis is zero (clearly $f$ cannot be
discontinuous as it is a solution of the homogeneous second--order
differential equation). So let us concentrate on the remaining cases and to
prove that then $a^{k}b^{l}[K=f]=0$.

\vspace{3mm}

\noindent To show this let us formulate the following conjecture:
\begin{equation}
a^{k}b^{l}[K]= \int_{C} d^{2}x d^{2}y\, \Delta_{C}(x;y)\left\{ \delta(x-y)
F^{(k,l)}[K](x) + B^{(k,l)}[K](x,y) \right\} \, ,
\label{B22}
\end{equation}
\noindent where both $F^{(k,l)}[K](x)$ and $B^{(k,l)}[K](x,y)$ are continuous
functions across the real--time axis  when $K=f$, i.e.
\begin{displaymath}
Disc_{x_{0}}F^{(k,l)}[f](x) = Disc_{x_{0}}B^{(k,l)}[f](x,y)=
Disc_{y_{0}}B^{(k,l)}[f](x,y)=0\, .
\end{displaymath}
\noindent To prove the conjecture (\ref{B22}) we shall use the mathematical
induction w.r.t. $k$. For $k=1$ the conjecture is true as
\begin{eqnarray}
ab^{l}[K] &=& \int_{C}d^{2}x d^{2}y\, \Delta_{C}(x;y) \left\{ l(l-1)
b^{l-2} \frac{\delta b}{\delta K(x)}\frac{\delta b}{\delta K(y)} + l\,
b^{l-1}\frac{\delta^{2} b}{\delta K(x) \delta K(y)} \right\} \nonumber \\
&=&  \int_{C} d^{2}x d^{2}y\, \Delta_{C}(x;y)\left\{ \delta(x-y)
F^{(1,l)}[K](x) + B^{(1,l)}[K](x,y) \right\}\, .
\label{B23}
\end{eqnarray}
\noindent From the structure of the $b$
term is obvious that both $F^{(1,l)}[f]$
and $B^{(1,l)}[f]$ are continuous across the real--time axis.

\vspace{3mm}

\noindent In the next induction step we shall assume
that the relation (\ref{B22})
is valid for $k=n$ and we should prove the validity for $k=n+1$.
In the latter case we may write
\begin{eqnarray}
a^{n+1}b^{l}[K] = a(a^{n}b^{l})[K] &=& \int_{C}d^{2}x d^{2}y\,
\Delta_{C}(x;y)
\frac{\delta^{2}}{\delta K(x) \delta K(y)}\, \int_{C} d^{2}z_{1}
d^{2}z_{2}\, \Delta_{C}(z_{1};z_{2})\nonumber \\ &\times & \left\{
B^{(n,l)}[K](z_{1},z_{2}) + \delta(z_{1}-z_{2}) F^{(n,l)}[K](z_{1})
\right\}\, .
\label{B24}
\end{eqnarray}
\noindent Because of the assumed property of $B^{(n,l)}$ and $F^{(n,l)}$
they may by written as
\begin{eqnarray}
B^{(n,l)}[K](x,y) &=& \sum_{p,q} f_{pq}(x,y) K^{p}(x)K^{q}(y) \,
P_{pq}[b]\nonumber \\
F^{(n,l)}[K](x) &=& \sum_{p} f_{p}(x) K^{p}(x)\, P_{p}[b]\, ,
\label{B25}
\end{eqnarray}
\noindent Where both $f_{pq}$ and $f_{p}$ are continuous across the
real--time axes,  $P_{pq}[b]$ and $ P_{p}[b]$ are some polynomials in $b$
(not in $K$).
Plugging (\ref{B25}) into (\ref{B24}) we obtain after some manipulations
that
\begin{equation}
a^{n+1}b^{l}[K] = \int_{C}d^{2}x d^{2}y\,  \Delta_{C}(x;y)\left\{
B^{(n+1,l)}[K](x,y) + \delta(x-y) F^{(n+1,l)}[K](x) \right\}\, ,
\label{B26}
\end{equation}
\noindent with  $B^{(n+1,l)}[f]$ and $F^{(n+1,l)}[f]$ fulfilling the
required conditions. For example $F^{(n+1,l)}[K]$ can be found to be
\begin{eqnarray*}
F^{(n+1,l)}[K](x) &=& \Delta_{11}(0) \, \sum_{p}f_{p}(x)
p(p-1)K^{p-2}(x)\, P_{p}[b] \\
&-& \frac{i}{2} \Delta_{11}(0) \, [3g^{2}K^{2}(x) + 6mg
K(x)]\,\frac{dP_{q}[b]}{db}\int_{C} d^{2}z_{1} \, \sum_{p}f_{p}(z_{1})
K^{p}(z_{1})\, .
\end{eqnarray*}
\noindent During the previous derivations we have used the fact that
\begin{equation}
\int_{C}d^{2}y\, \Delta_{C}(x;y) f^{p}(y) = i\int_{-\infty}^{\infty}
d^{2}y\, G_{R}(x,y) f^{p}(y) = z(x) \, ,
\label{B34}
\end{equation}
\noindent where $z(x)$ is a continuous across $x_{0}$. This is obvious as
$z(x)$ solves equation $(\partial^{2} + m^{2}) z(x) = f^{p}(x)$ with the
boundary conditions $z(0, x_{1}) =0$ and $dz(0,x_{1})/dx_{0} =0$ (see
for example \cite{Vladimirov}). The second important trick which we have
used was that
\begin{equation}
\int_{C}d^{2}xd^{2}y\, \Delta_{C}(x,y) \, \delta (x-y) \ldots \; \,=
\Delta_{11}(0) \int_{C}d^{2}x \, \ldots \, .
\label{B35}
\end{equation}
\noindent This closes the proof of our conjecture (\ref{B22}). As a result
we may write
\begin{equation}
a^{k}b^{l}[f] = \Delta_{11}(0) \, \int_{C}d^{2}x\, F^{(k,l)}[f](x) +
\int_{C}d^{2}x d^{2}y \, \Delta_{C}(x;y)\, B^{(k,l)}[f](x,y) =0\, .
\label{B27}
\end{equation}
In the previous we have applied both the fact that the contour
integral over the continuous function across the real--time axis
is zero and  the identity
\begin{equation}
\Delta_{11} - \Delta_{12} - \Delta_{21} + \Delta_{22} = 0 \, ,
\lab{eq11}
\end{equation}
\noindent which renders the double--contour integration zero\footnote{As
the identity (\ref{eq11}) is only based on the fact
that $i\Delta_{11}(x) = (i\Delta_{22}(x))^{*} =
\theta(x_{0})i\Delta_{21}(x) + \theta(-x_{0})i\Delta_{12}(x)$, it
holds both for $\Delta_{C}(x;y)$ and $\Delta_{C}(x;y,T)$. Note that the
latter automatically implies that
\begin{displaymath}
\Delta_{11}^{2} - \Delta_{12}^{2} -
\Delta_{21}^{2} + \Delta_{22}^{2}
= (\Delta_{11} + \Delta_{22})^{2} - (\Delta_{12} + \Delta_{21})^{2} -
2\Delta_{11}\Delta_{22} + 2\Delta_{12}\Delta_{21} = 0\, ,
\end{displaymath}
\noindent and so $\int_{C} d^{2}x d^{2}y\, \Delta(x;y) \Delta(x;y)\,
F(x,y) = 0$ provided $Disc_{x_{0}}F(x,y) = Disc_{y_{0}}F(x,y) = 0$.}.
As a byproduct we obtain that only the first term in (\ref{B1}) survives,
i.e.  $D[f,\hbar \rightarrow 0] = f$.

\vspace{3mm}

\noindent One of the main virtue of our derivation is that it
immediately extends to finite temperatures because all the steps
in the above proof may be repeated almost word by word. This
mainly goes into account of equations (\ref{B34}), (\ref{B35}) and
(\ref{eq11}) (and the arguments mentioned therein) which retain
their validity even at the finite temperature level.

\section*{Appendix B}

\noindent In this appendix we aim to prove the recursion relation
(\ref{44}).  Let us first observe that for $n=2$ we immediately obtain
\bea &&P_{2}(x) = \int_{-\infty}^{\infty} d^{2}y \, G_{R}(x,y)
\,\lf[\frac{3}{2}mg P_{1}(y)\,P_{1}(y) + \frac{1}{2}g^{2}
P_{1}(y)\,P_{1}(y)\,P_{1}(y)\ri]\,.  \eea
\noindent with $P_{1} = K$. Similarly, for $n=3$ we may write
\bea &&P_{3}(x) = i \int_{-\infty}^{\infty} d^{2}y \,G_{R}(x,y) \,
a\lf\{\lf[ \frac{3}{2}mg K^{2}(y) + \frac{1}{2}g^{2}
K^{3}(y)\ri]b\ri\}\non \\ &&\non \\ &&\non \\ &&\mbox{\hspace{5mm}}=
\int_{-\infty}^{\infty}d^{2}y \,
G_{R}(x,y)\,\lf[\frac{3}{2}mg2\,P_{1}(y)\,P_{2}(y) +
\frac{g^{2}}{2}3\,P_{1}(y)\,P_{1}(y)\,P_{2}(y)\ri]\, .  \eea
\noindent During the previous derivation we have took advantage of
the relation $\Delta(x,y) = -\Delta(y,x)$. No assumption about the
actual behavior of $K(x)$ was made at this stage. As a result we
may conjecture that for the general $n$ we have
\bea P_{n}(x) &=& \int_{-\infty}^{\infty}d^{2}y \, G_{R}(x,y) \lf[
\frac{3}{2}mg \,\sum_{i+j=n}P_{i}(y)\,P_{j}(y) + \frac{1}{2}g^{2}\,
\sum_{i+j+k=n+1}P_{i}(y)\,P_{j}(y)\,P_{k}(y)\ri], \, n\ge 2.
\lab{A33} \eea
\noindent To prove this conjecture we shall use the mathematical
induction. Let us assume that Eq.(\ref{A33}) holds for $n$. So for
$P_{n+1}(x)$ we may directly write
\bea &&P_{n+1}(x) = a\lf[P_{n}(x) b\ri]\, \frac{1}{n^{2}}\non \\
&&\non \\ &&\mbox{\hspace{5mm}}= \frac{1}{n!}
\int_{-\infty}^{\infty}d^{2}y \, G_{R}(x,y)  \,\lf\{\frac{3}{2}mg\,
a\lf[\sum_{i+j=n} P_{i}(y)\,P_{j}(y) b\ri]\frac{1}{n^{2}} +\,
\frac{1}{2}g^{2} a\lf[\sum_{i+j+k=n+1}
P_{i}(y)\,P_{j}(y)\,P_{k}(y)b\ri]\frac{1}{n^{2}}\ri\}\non \\ &&\non \\
&&\non \\ &&\mbox{\hspace{5mm}}= \frac{1}{n!}
\int_{-\infty}^{\infty}d^{2}y \, G_{R}(x,y)  \, \lf[\frac{3}{2}mg\,
\sum_{i+j=n+1} P_{i}(y)P_{j}(y)\,+ \frac{1}{2}g^{2}\, \sum_{i+j+k=n+2}
P_{i}(y)\,P_{j}(y)\,P_{k}(y)\ri]\, .  \lab{A31} \eea
\noindent This proves our conjecture.

\section*{Appendix C}

\noindent We calculate here the following integral:
\bea I_n\,=\,\int_C d^2z \,\De_C(x-z) \, f^n(z) \eea
with $f(z) \, =\, A\, e^{-m \ga (z_1 - u z_0)}$.  We get
\bea I_n\,&=&\,
A^n\int_{-\infty}^{\infty}dz_0\int_{-\infty}^{\infty}dz_1
\lf\{\De_{11}(x-z) - \De_{12}(x-z)\ri\}\, e^{-nm\ga(z_1 - u z_0)} \eea
By performing the change of variables $x-z=y$ and using
Eq.(\ref{pjordan}) for $\De_{11}(x-z) - \De_{12}(x-z) = \te(x_0-z_0)
[\phi(x),\phi(z)]$, we get
\bea I_n &=& -\frac{i}{2} A^n e^{-n m\ga(x_1 - u x_0)}
\int_{0}^{\infty}dy_0\int_{-y_0}^{y_0}dy_1 \, J_0[m\sqrt{y_0^2-y_1^2}]
\, e^{n m\ga(y_1 - u y_0)} \non \\ &&\non \\ &=& - \frac{i}{2} A^n e^{-n
m\ga(x_1 - u x_0)} \int_{0}^{\infty}dy_0 e^{-n m\ga u y_0}
y_0\int_{-1}^{1}dw  \, J_0[m y_0\sqrt{1 - w^2}] \, e^{n m\ga y_0 w }
\eea
where we used $w=y_1/y_0$. We then have (see\cite{GR1} formula 6.616
n.5 after analytic continuation):
\bea I_n &=& - i A^n \frac{e^{-n m\ga(x_1 - u x_0)}}{m \sqrt{n^2\ga^2
-1}} \int_{0}^{\infty}dy_0 \,  e^{-n m\ga u y_0} \, \sinh(m
y_0\sqrt{n^2\ga^2 -1} ) \non \\ && \non \\ &=& - i A^n \frac{e^{- n
m\ga(x_1 - u x_0)}}{m^2 (n^2\ga^2 -1)} \int_{0}^{\infty}dy_0' \,
e^{-\frac{n \ga u }{\sqrt{n^2\ga^2 -1}}y_0'} \sinh(y_0' )\non \\ \eea
\noindent The last integral is tabulated (see\cite{GR1}, 3.541 n.1)
with the result
\bea \int_{0}^{\infty}dx \,  e^{-\mu x} \sinh(x) &=& \frac{1}{4}
B\left(\frac{\mu}{2} -\frac{1}{2}, 2\right) \, =\, \frac{1}{4}
\frac{\Ga(\frac{\mu}{2} - \frac{1}{2})\Ga(2)} {\Ga(\frac{\mu}{2} +
\frac{3}{2})} \, =\, \frac{1}{\mu^2-1}\, , \eea
\noindent where $\mu \equiv n \ga u /\sqrt{n^2\ga^2 -1}$.  The final
result is then
\bea I_n \, &=&\, - i A^n \frac{e^{- n m\ga(x_1 - u x_0)}}{m^2 (n^2\ga^2
-1)} \frac{1}{\mu^2-1} \,=\,  i A^n \frac{e^{- n m\ga(x_1 - u
x_0)}}{m^2 (n^2\ga^2 -1)} \frac{n^2\ga^2 -1}{n^2\ga^2(1-u^2) -1} \,
=\, i A^n \frac{e^{- n m\ga(x_1 - u x_0)}}{m^2 (n^2 -1)} \eea
So $\int_{C}d^{2}z \, \Delta_{C}(x-z)f^{n}(z) \propto f^{n}(x)$
provided $f$ is an exponential solution of the Klein--Gordon
equation (\ref{cpt3b}). On the other hand an exponential function
is the only one which has this property for all $n$. To see this
let us reverse the former integral equation, i.e.
\begin{equation}
(\partial^{2} + m^{2})f^{n} = (\partial_{+}\partial_{-} +m^{2})f^{n} =
\eta f^{n} \, ,
\end{equation}
\noindent where $\eta$ is a proportionality constant and
$\partial_{\pm}= \partial /\partial x^{\pm}$  with $x^{\pm} =
\frac{1}{\sqrt{2}}(x^{0} \pm x^{1})$. As a result we get for $n>1$
the non--linear differential equation
\begin{displaymath}
\partial_{+}f \partial_{-}f =\xi^2 f^2\, ,
\end{displaymath}
with $\xi = \sqrt{m^{2}/(n-1) + \eta /n(n-1)}$ which can be
equivalently written as
\begin{displaymath}
(\partial_+ f + \xi f)(\partial_- f + \xi f) =0\, .
\end{displaymath}
The latter has clearly only exponential solutions.

\section*{Appendix D}

\noindent In this appendix we show how the classical kink solutions emerge in
the small coupling (large mass) limit. Let us put $K \rightarrow K/g $. Then,
as already done in Section III.B, we may argue that since
${\cal{L}}_{ct}$ is of order ${\cal{O}}(g^{2})$ it automatically
decouples in the following reasonings. As a result we obtain:

\begin{eqnarray}
D[K](x) &=& \frac{K(x)}{g}\, \exp[\hbar g^{2}\, a]\,
\exp\left[\frac{1} {\hbar g^{2}}\, {\cal{B}}\right]
\nonumber \\ C[K](x) &=&  \hbar g \, \int_{C} d^{2}y \,
\Delta_{C}(x,y) \, \frac{\delta}{\delta K(y)}\, \exp[\hbar
g^{2}\, a] \, \exp\left[\frac{1}{\hbar g^{2}} \, {\cal{B}}
\right]\, , \label{dB2}
\end{eqnarray}
\noindent with $ {\cal{B}} = - \frac{i}{2}\, \int_{C} d^{2}z \,
\left[ \frac{K^{4}(z)}{4} + \omega_{0}K^{3}(z)  \right]\, .
$
\noindent Keeping only the leading terms in the $g \rightarrow 0$
limit (i.e. terms of order  $g^{-1}$) we obtain
\begin{eqnarray}
D[K](x, g \rightarrow 0) &=&  \frac{K(x)}{g} \,Res_{g \rightarrow
0} \left(  \frac{1}{g} \, \exp[\hbar g^{2} a] \,
\exp\left[ \frac{{\cal{B}}}{\hbar g^{2}} \right] \right)  =
\frac{K(x)}{g} \, \sum_{n=0} \frac{1}{(n!)^{2}}\,
a^{n}{\cal{B}}^{n} = \frac{K(x)}{g} \, , \label{dB3}
\end{eqnarray}
\noindent where the passage between second and third identity was done
by means of the result in Appendix A. Analogously, for the $C[K]$
term we arrive at
\begin{eqnarray}
C[K](x, g \rightarrow 0) &=&  \frac{\hbar}{g} \,\int_{C}d^{2}y\,
\Delta_{C}(x,y)\, \frac{\delta}{\delta K(y)}  \, Res_{g
\rightarrow 0}\left( g \,\exp[\hbar g^{2} a]\, \exp
\left[\frac{1}{\hbar g^{2}}{\cal{B}} \right] \right)\nonumber \\
&=&  \frac{i}{g} \int_{- \infty}^{\infty} d^{2}y\, G_{R}(x,y) \,
\frac{\delta}{\delta K(y)} \, \sum_{n=0}^{\infty}
\frac{1}{n!(n+1)!}\,  a^{n}{\cal{B}}^{n+1} \, . \label{dB4}
\end{eqnarray}
\noindent Let us remark that $\hbar$ completely disappeared both  from
(\ref{dB3}) and (\ref{dB4}).

\vspace{3mm}

\noindent Similarly as in the case of the $\hbar$ limit we may
write
\begin{eqnarray}
\langle \psi^{f}(x) \rangle_{g \rightarrow 0} &=& v  + \frac{f}{g}
 + \frac{i}{g} \, \int_{- \infty}^{\infty} d^{2}y \, G_{R}(x,y) \,
\frac{\delta}{\delta K(y)} \,  \sum_{n=0}^{\infty}
\frac{1}{n!(n+1)!} \, a^{n}{\cal{B}}^{n+1}|_{K=f}\nonumber \\ &=&
v + \frac{1}{g}\,\sum_{n=1}^{\infty} R_{n}[K](x)|_{K=f}\, ,
\label{dB5}
\end{eqnarray}
\noindent where
\begin{eqnarray}
R_{1}(x) &=&  K(x)\nonumber \\ R_{n}(x) &=&
\frac{1}{[(n-2)!]^{2}} \, \int_{-\infty}^{\infty}d^{2}y  \,
G_{R}(x,y) \, a^{n-2}  \left\{ \left[\frac{3}{2}
\omega_{0}K^{2}(y) + \frac{1}{2} K^{3}(y) \right] {\cal{B}}^{n-2}
\right\} \, ; \;\;\; n \ge 2\, . \label{dB6}
\end{eqnarray}
\noindent It is seen comparing Eg.(\ref{dB6}) with Eq.(\ref{pj1})
that the following recurrence relation holds:
\begin{equation}
R_{n}(x) = - \int_{- \infty}^{\infty} d^{2}y\, G_{R}(x,y)\, \left[
\frac{3\omega_{0}}{2}\, \sum_{i+j = n} R_{i}(y)R_{j}(y) +
\frac{1}{2}\, \sum_{i+j+k = n+1} R_{i}(y)R_{j}(y)R_{k}(y)\right] \,
; \;\;\; n \ge 2 \, . \label{dB7}
\end{equation}
\noindent The ``diagonalized'' recurrence relation reads
\begin{eqnarray}
S_{1}(x) &=& K(x) \nonumber \\   S_{2}(x) &=&
\int_{-\infty}^{\infty} d^{2}y \,G_{R}(x,y)\,
\frac{3\omega_{0}}{2}\, \sum_{i+j = 2} S_{i}(y)S_{j}(y) \nonumber
\\ S_{n}(x) &=&  \int_{-\infty}^{\infty} d^{2}y \,
G_{R}(x,y)\,\left[ \frac{3 \omega_{0}}{2}  \, \sum_{i+j = n}
S_{i}(y) S_{j}(y) + \frac{1}{2} \sum_{i+j+k = n} S_{i}(y)S_{j}(y)
S_{k}(y) \right]\, ; \;\;\; n \ge 3\, . \label{dB8}
\end{eqnarray}
\noindent This is, as before,  the functional equation of
Cauchy--Marley's type.  The corresponding analytical solution has
the form $S_{n}(x)  \propto (S_{1}(x))^{n}$, with $S_{1}(x) =
f(x)$ being an exponential, Fourier non--transformable and $g$
independent solution of the dynamical equation (\ref{cpt3b}). So
\begin{displaymath}
S_{n}(x) = {\tilde{A}}_{n}f^{n}(x) = {\tilde{A}}_{n}\, e^{-
\omega_{0} n \gamma  (x_{1} - x_{0}u)} \, .
\end{displaymath}
Incorporating this solution in the recurrence relation
(\ref{dB8}), we are led to the conclusion that
\begin{equation}
{\tilde{A}_{n}} = \frac{1}{(n^{2}-1)}\, \left\{ \frac{3}{2
\omega_{0}} \, \sum_{i+j = n} {\tilde{A}}_{i} {\tilde{A}}_{j} +
\frac{1}{2 \omega_{0}^{2}}\, \sum_{i+j+k = n}
{\tilde{A}}_{i}{\tilde{A}}_{j}{\tilde{A}}_{k} \right\} \, .
\end{equation}
\noindent By analogy with the calculation of Section III.B  it is
evident that  the fundamental solution of this recurrence equation
has the form ${\tilde{A}}_{n} = 2 \omega_{0} \, \left( \frac{s}{2
\omega_{0}} \right)^{n}$ where $s$ is a real constant. The final
form for the order parameter in the small coupling limit may be
written as
\begin{equation}
\langle \psi^{f}(x) \rangle_{g \rightarrow 0} = v + \frac{2
\omega_{0}}{g} \, \sum_{n=1}^{\infty} \left( \frac{s f(x)}{ 2
\omega_{0}} \right)^{n} \, . \label{dB10}
\end{equation}
\noindent Recalling that $v = \pm \omega_{0}/g$ and choosing $s=
\pm 2\omega_{0}\, e^{\omega_{0} \gamma a}$ we can resum the series
(\ref{dB10}) with the result:
\begin{equation}
\langle \psi^{f}(x) \rangle_{g \rightarrow 0} =  v \, \mbox{th}
\left[ \frac{\omega_{0}}{2} \gamma ((x_{1} -a) - ux_{0} )\right]
\, , \label{dB11}
\end{equation}
\noindent for $v$ positive and $s$ negative, and
\begin{equation}
\langle \psi^{f}(x) \rangle_{g \rightarrow 0} = - v \, \mbox{th}
\left[ \frac{\omega_{0}}{2} \gamma ((x_{1} -a) - ux_{0} )\right]
\, , \label{dB12}
\end{equation}
\noindent for both $v$ and $s$ negative. If $s$ were positive, the
expression (\ref{dB10}) for the order parameter becomes   the
non--solitonic one with $\mbox{cth}[\ldots]$ instead of
$\mbox{th}[\ldots]$. An analogous analysis can be done
for $f = e^{\omega_{0}\gamma (x_{1} - x_{0}u)}$.

\end{document}